\documentclass[a4paper,11pt]{article}
\usepackage{jheppub} 
\usepackage[T1]{fontenc} 
\RequirePackage[displaymath]{lineno} 
\usepackage[perpage,symbol]{footmisc}
\usepackage{epsfig}
\usepackage{siunitx}
\usepackage{graphicx}
\usepackage{dcolumn}
\usepackage{bm}
\usepackage{ltablex,booktabs}
\usepackage{overpic}
\usepackage{subfigure}
\usepackage{float}
\usepackage{color}
\usepackage{amsmath}
\usepackage{mathcomp}
\usepackage{mathrsfs}
\usepackage{multirow}
\usepackage{rotating}
\usepackage{amssymb}
\usepackage{gensymb}
\usepackage{amsmath}
\usepackage{tabularx}
\usepackage{epstopdf}
\usepackage{url}
\usepackage{tikz-feynman}
\usepackage{subcaption} 
\usepackage{hyperref}
\usepackage{subcaption}
\newcommand{\BESIIIorcid}[1]{\href{https://orcid.org/#1}{\hspace*{0.1em}\raisebox{-0.45ex}{\includegraphics[width=1em]{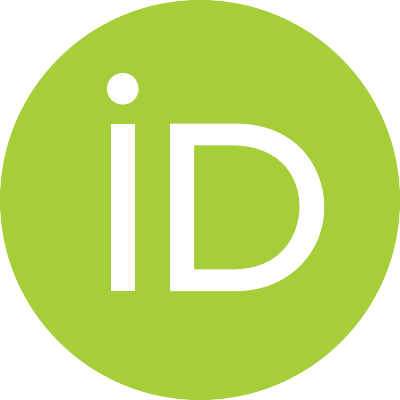}}}} 

\title{\textbf{Measurement of the relative phase between the strong and electromagnetic amplitudes in $\boldsymbol{J/\psi \to \Sigma^{+} \bar{\Sigma}^{-}}$}}
\collaborationImg{\includegraphics[width=0.15\textwidth,angle=90]{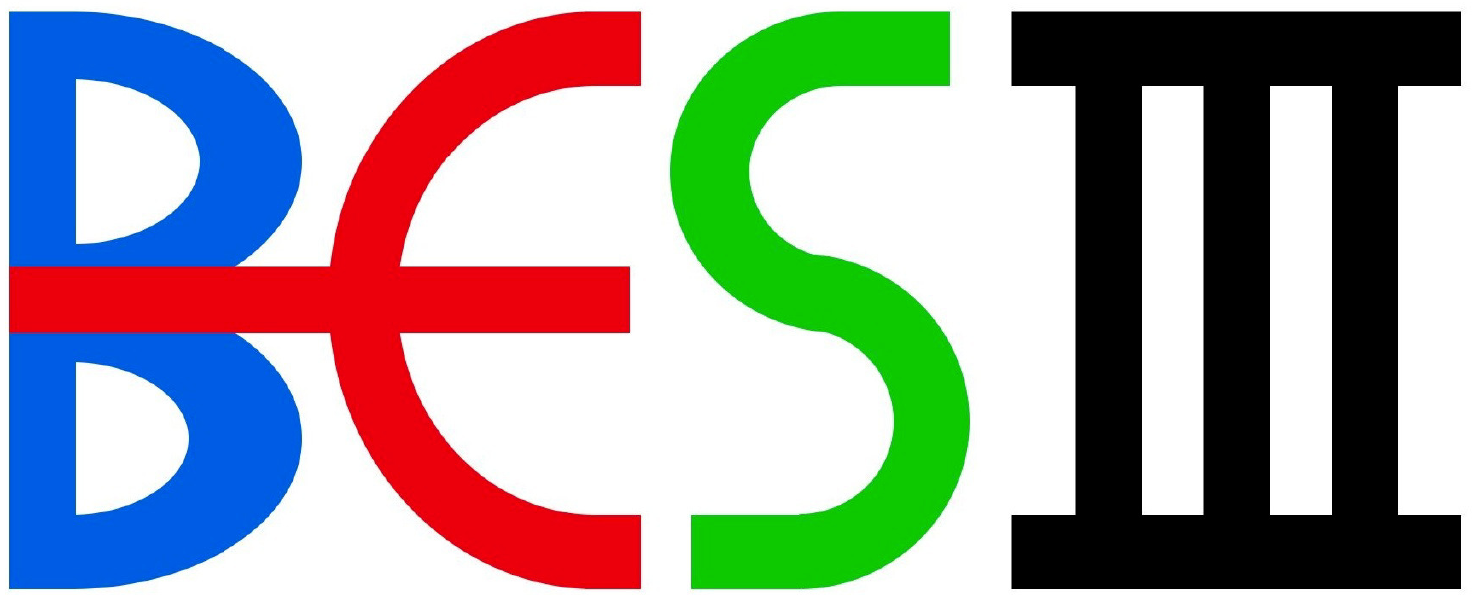}}
\collaboration{The BESIII Collaboration}

\date{\today}
\abstract{Using 284 pb$^{-1}$ of $e^{+}e^{-}$ collision data collected with the BESIII detector at center-of-mass energies between 3.00 and 3.12 GeV, the relative phase between the strong and electromagnetic interaction amplitudes in the process $J/\psi \to \Sigma^+ \bar{\Sigma}^-$ is determined.
The relative phase is measured to be $(107.9 \pm 24.9)^{\circ}$ or  $(-107.6 \pm 24.3)^{\circ}$, where the uncertainty includes both statistical and systematic sources. This is the first experimental determination of the strong-electromagnetic phase difference in $J/\psi$ decays to a hyperon-antihyperon pair.
The branching fraction $\mathcal{B}_{J/\psi \to \Sigma^+ \bar{\Sigma}^-}$ is measured to be $(1.14 \pm 0.04)\times 10^{-3}$ for the positive phase solution or 
 $(1.19 \pm 0.04) \times 10^{-3}$ for the negative phase solution. These results provide important input for testing perturbative QCD predictions and studies of SU(3) flavor symmetry.}

\arxivnumber{}
\begin{document}
\begin{sloppypar}
\maketitle
\flushbottom

\section{Introduction}

\begin{figure*}[htbp]
	\centering
	
	\includegraphics[width=4.5cm,height=1.8cm]{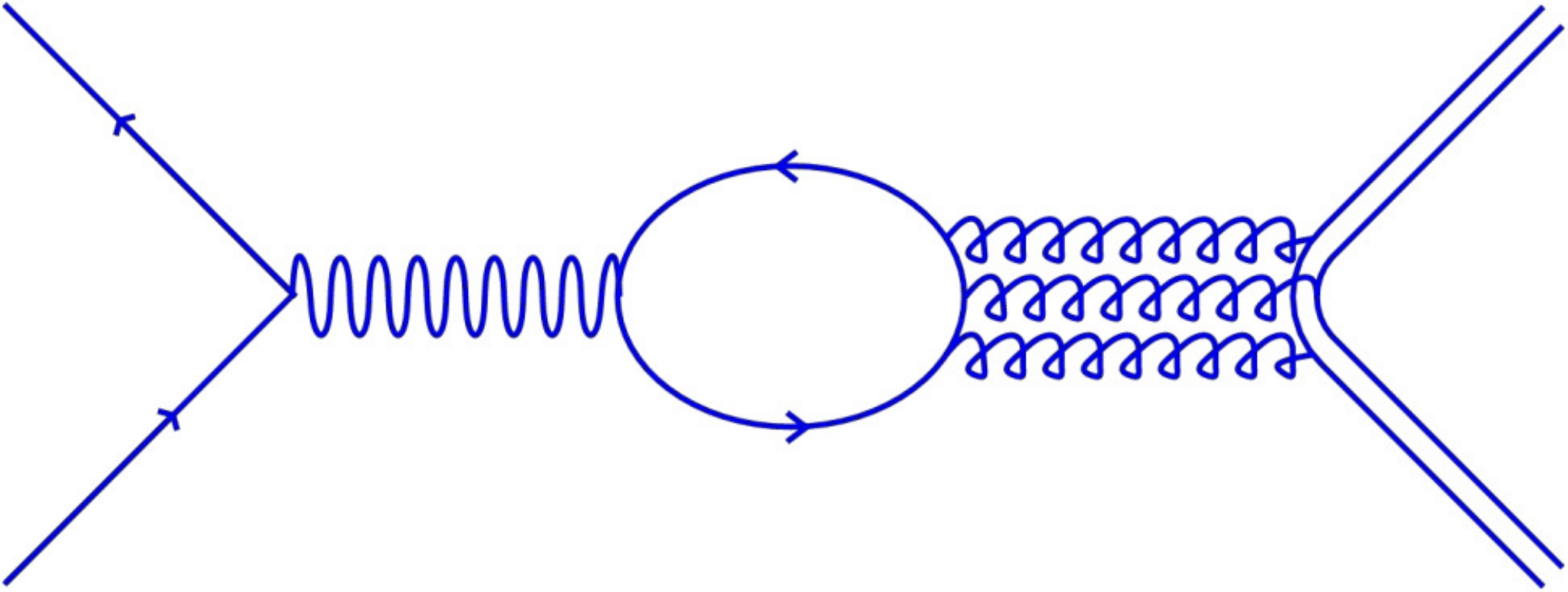}%
	\put(-64,45){\makebox[0pt]{$\mathcal{A}_{3g}$}}%
	\put(-64,0){\makebox[0pt]{(a)}}%
	\hskip 5mm
	\includegraphics[width=4.5cm,height=1.8cm]{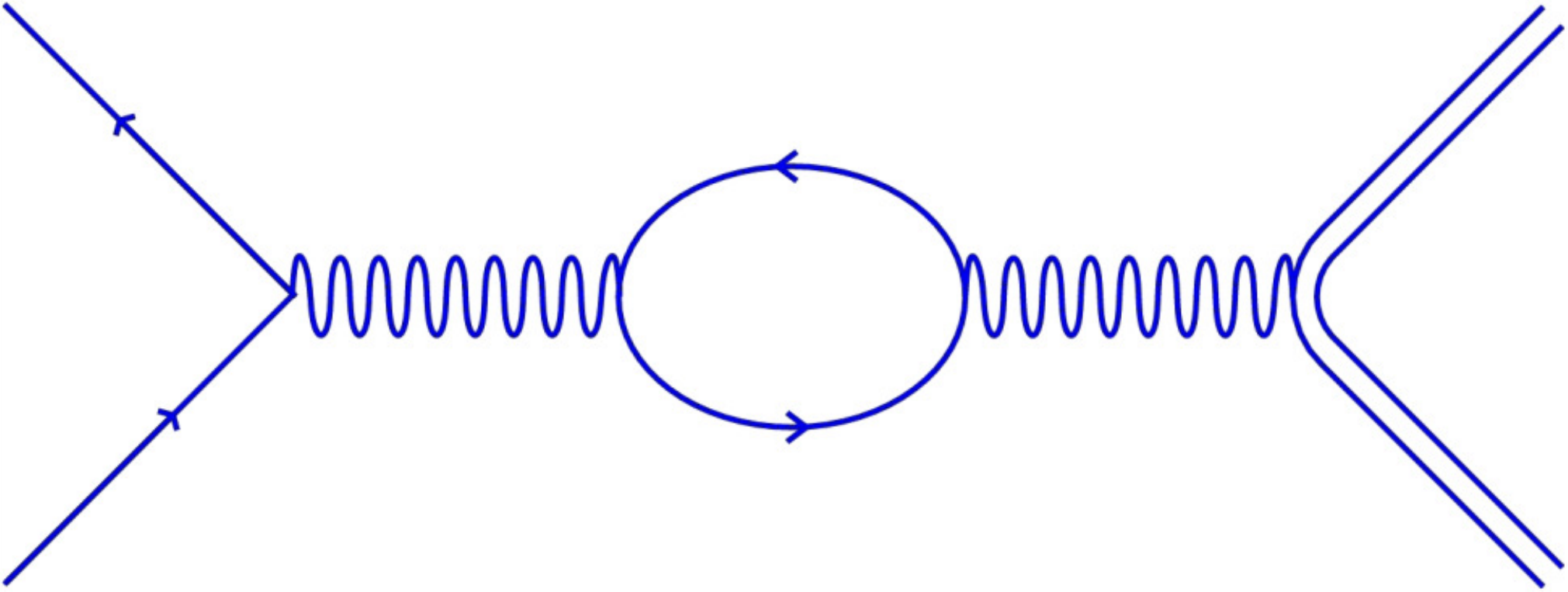}%
	\put(-64,45){\makebox[0pt]{$\mathcal{A}_{\gamma}$}}%
	\put(-64,0){\makebox[0pt]{(b)}}%
	\hskip 5mm
	\includegraphics[width=4.5cm,height=1.8cm]{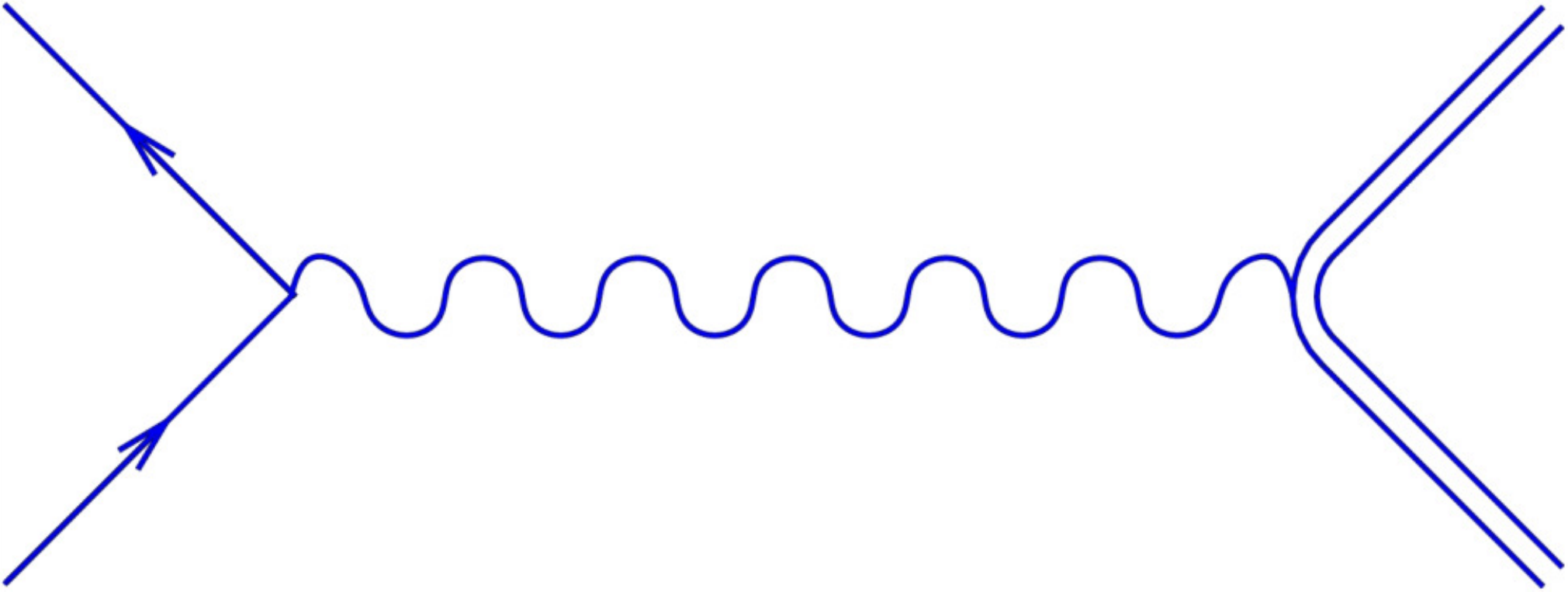}%
	\put(-64,45){\makebox[0pt]{$\mathcal{A}_{C}$}}%
	\put(-64,0){\makebox[0pt]{(c)}}%
	
	\caption{Feynman diagrams contributing to the process $e^+ e^- \to \text{hadrons}$:
		(a) the purely strong resonant amplitude mediated by 3 gluons;
		(b) the purely EM resonant amplitude mediated by a virtual photon; and
	    (c) the non-resonant (continuum) amplitude also proceeding via a virtual photon.}
	\label{fig:feyman}
\end{figure*}

Since the discovery of the $J/\psi$ particle in 1974, extensive investigations have characterized its properties, production mechanisms, and decay dynamics.  These studies have revealed that the primary decay channels of the $J/\psi$ are predominantly mediated by strong and electromagnetic (EM) interactions. Previous studies have shown that the phase difference between strong and EM interactions is a measurable parameter, providing key insights into the dynamic interplay between these two interactions~\cite{SLAC-SP-017:1974ind, Freund:1975pn, MARK-III:1988crp,Wang:2002ef,Czyz:2009vj, Zhou:2017yod,Guo:2022gkg}.
In the vicinity of the $J/\psi$ meson, the annihilation of $e^{+}e^{-}$ into hadronic final states can be parameterized in terms of three amplitudes~\cite{BESIII:2018wid}: a purely strong decay of the $J/\psi$ (mediated by 3 gluons) $\mathcal{A}_{3g}$, a purely EM decay (mediated by a virtual photon) $\mathcal{A}_{\gamma}$, and a continuum process (i.e. an EM amplitude mediated by a virtual photon without an intermediate $J/\psi$ resonance)  $\mathcal{A}_{C}$, as shown in figure~\ref{fig:feyman}.
The relative phase between the strong and electromagnetic amplitudes of quarkonium decays is a basic parameter that provides insight into the dynamics of quarkonium decays. The phase term explicitly appears in the total cross section, which can be expressed as:
\begin{equation}
	\sigma_{\text{hadron}} \propto \left|\left[\mathcal{A}_{3g}(W)e^{i\Phi_{3g,\gamma}} + \mathcal{A}_{\gamma}(W)\right]e^{i\Phi_{\gamma,C}} + \mathcal{A}_C(W)\right|^2,
\end{equation}
where $\Phi_{3g,\gamma}$ is the phase difference between $\mathcal{A}_{3g}$ and $\mathcal{A}_{\gamma}$, and $\Phi_{\gamma,C}$ denotes the phase difference between the EM and continuum amplitudes, and $W$ is the center-of-mass (c.m.) energy. Since the electromagnetic continuum amplitude is purely real at leading order, and experimental studies of processes like $ e^+e^- \to \mu^+ \mu^-$ and $e^+e^- \to \eta \pi^+ \pi^-$~\cite{BESIII:2018wid} have confirmed $\Phi_{\gamma,C} = 0^{\circ}$, the  cross section can be simplified to:
\begin{equation}
            \sigma_{\text{hadron}}\propto |\mathcal{A}_{3g}(W)e^{i\Phi_{3g,\gamma}}+\mathcal{A}_{\gamma}(W)+\mathcal{A}_{C}(W)|^2.
\end{equation}

In perturbative Quantum Chromodynamics (pQCD), $\Phi_{3g,\gamma}$ is expected to be $0^{\circ}$ or $180^{\circ}$ at the lowest order of approximation, indicating full interference between the two amplitudes~\cite{Chernyak:1984bm, Brodsky:1987bb}.  
In contrast to this prediction, model-dependent analyses incorporating SU(3) flavor symmetry and symmetry-breaking reveal that $\Phi_{3g,\gamma}$ in $J/\psi$ two-body decays (including spin-parity combinations such as $1^-0^-,$~\cite{MARK-III:1988crp, DM2:1988bfq} $ 0^-0^-,$~\cite{LopezCastro:1994xw, Suzuki:1999nb, Kopke:1988cs}  $ 1^-1^-$~\cite{Kopke:1988cs}, $1^+0^-$~\cite{Suzuki:2000yq} and baryon-antibaryon~\cite{LopezCastro:1994xw, Baldini:1998en,Zhu:2015bha}) is found to be close to 90$^\circ$. 
Similarly,  model-dependent studies indicate that  $\psi(2S)$ decays to pseudoscalar-pseudoscalar meson pairs~\cite{CLEO:2006udi} exhibit phase alignment comparable to $J/\psi$ channels, whereas analyses of pseudoscalar-vector final states~\cite{Suzuki:2000yq, Chen:1998ma} reveal that $\Phi_{3g,\gamma}$ is around $0^{\circ}$. 
Verification of these models through model-independent experiments is essential, which can be achieved by measuring the cross-section lineshape around the resonance.
Notably, BESIII measured $\Phi_{3g,\gamma}$ for the $J/\psi$ in a model-independent analysis of $e^+e^- \to 5\pi$ to be $(84.9\pm3.6)^\circ$ or $(-84.7\pm3.1)^\circ$~\cite{BESIII:2018wid}, aligning with the $90^\circ$ trend.  
Precise experimental measurements can provide crucial constraints and validation for theoretical QCD calculations. In particular, as demonstrated in Ref.~\cite{Guo:2022gkg}, the interference effect associated with this relative phase is crucial for measurements of $J/\psi$ and $\psi(2S)$ branching fractions aiming at percent-level or better precision.

In this work, we present the first model-independent measurement of the relative phase $\Phi_{3g,\gamma}$ in $ e^+e^- \to \Sigma^+ \bar{\Sigma}^-$ at 27 c.m. energy points in the $J/\psi$ resonance region ($\sqrt{s}$ = 3.00-3.12 GeV) with an integrated luminosity of  284~pb$^{-1}$. The determination of this phase provides critical tests of pQCD predictions and SU(3) flavor symmetry in hyperon pair production.

\section{BESIII detector and Monte Carlo simulation}
\label{sec:detector_dataset}

The BESIII detector~\cite{BESIII:2009fln} records symmetric $e^+e^-$ collisions
provided by the BEPCII storage ring~\cite{Yu:2016cof}
in the center-of-mass energy range from 1.84 to 4.95~GeV,
with a peak luminosity of $1.1 \times 10^{33}\;\text{cm}^{-2}\text{s}^{-1}$
achieved at $\sqrt{s} = 3.773\;\text{GeV}$.
Large data samples have been collected in this energy region.
The cylindrical core of the BESIII detector covers 93\% of the full solid angle and consists of a helium-based
multilayer drift chamber~(MDC), a time-of-flight
system~(TOF), and a CsI(Tl) electromagnetic calorimeter~(EMC),
which are all enclosed in a superconducting solenoidal magnet
providing a 1.0~T magnetic field~\cite{Huang:2022wuo}.
The magnetic field was 0.9~T in 2012, which affects 35.4\% of the total integrated luminosity used in this analysis.
The solenoid is surrounded by an octagonal flux-return yoke made of steel, interleaved with resistive-plate-counter muon-identification modules.
The charged-particle momentum resolution at $1~{\rm GeV}/c$ is
$0.5\%$, and the
${\rm d}E/{\rm d}x$
resolution is $6\%$ for electrons
from Bhabha scattering. The EMC measures photon energies with a
resolution of $2.5\%$ ($5\%$) at $1$~GeV in the barrel (end cap)
region. The time resolution in the plastic scintillator TOF barrel region is 68~ps, while
that in the end cap region was 110~ps. The end cap TOF
system was upgraded in 2015 using multigap resistive plate chamber
technology, providing a time resolution of
60~ps~\cite{Li:2017jpg,Guo:2017sjt,Cao:2020ibk},
which benefits 11.0\% of the data used in this analysis.

The analysis uses data samples collected in 2012, 2015 and 2018 at c.m. energies between 3.000 and 3.1199 GeV.  The energies and integrated luminosities are  summarized in Table~\ref{table:dataset_and_obs_cs}.
During the 2012 energy scan experiment, the c.m.\ energies were measured by the Beam Energy Measurement System (BEMS)~\cite{BESIII:2017lkp}, in which photons from a CO$_2$ laser are Compton back-scattered off the electron beam and detected by a high-purity Germanium detector~\cite{Abakumova:2011rp}. As reported in Ref.~\cite{BESIII:2018wid}, the fitted value of the $J/\psi$ mass, $M_{\rm {fit}},$ is higher than its world average value $M_{{J/\psi}}$~\cite{ParticleDataGroup:2024cfk}, which indicates an offset in the absolute energy calibration for the BEMS. Therefore, the c.m.\ energies of Monte Carlo (MC) simulated event samples are corrected by shifting the absolute values with $\Delta M = (0.548 \pm 0.03)$ MeV, where $\Delta M = M_{\rm {fit}}- M_{{J/\psi}}$~\cite{BESIII:2018wid}. 
The integrated luminosity measurements are derived from well-measured Quantum Electrodynamics (QED) processes, specifically through the processes $e^{+}e^{-}\to \gamma_{\rm ISR} e^{+}e^{-}$ and $e^{+}e^{-}\to \gamma_{\rm ISR} \gamma \gamma$~\cite{BESIII:2017lkp}.

The optimization of the event selection criteria, estimation of background contamination, and evaluation of detection efficiencies are performed using dedicated MC samples for both signal and background processes. The MC samples are generated  using a {\sc Geant4}-based~\cite{GEANT4:2002zbu} package, which includes a description of the detector geometry and the detector response. 
The simulation models the beam-energy spread and initial-state radiation (ISR) in the $e^{+}e^{-}$ annihilations with the 
 {\sc ConExc}~\cite{Ping:2013jka}  generator.  An inclusive MC sample of $J/\psi$  decays is used to explore possible hadronic backgrounds.  In  this sample, the known decay modes are generated with {\sc EvtGen}~\cite{Lange:2001uf, Ping:2008zz} incorporating the branching fractions from the Particle Data Group (PDG)~\cite{ParticleDataGroup:2024cfk}, and the remaining unknown decays are generated using the {\sc LundCharm} generator model~\cite{Yang:2014vra}. Final state radiation from charged final state particles is incorporated using the {\sc Photos} package~\cite{Richter-Was:1992hxq}. Backgrounds coming from the QED background processes $e^+e^- \to \gamma_{\rm ISR} l^+l^- (l = e, \mu)$ and $e^+e^- \to \gamma_{\rm {ISR}} \gamma \gamma$ are generated with the generator {\sc Babayaga}~\cite{Balossini:2006wc, Balossini:2008xr}, while  {\sc Bestwogam}~\cite{Nova:1990prh}  is used for two-photon processes. 
 
 The process $e^{+}e^{-} \to \Sigma^{+} \bar{\Sigma}^{-}$ exhibits an angular distribution characterized by $1+\alpha_\theta \cdot \rm{cos}^{2}\theta_{\Sigma^+}$,  where $\alpha_\theta$ denotes the anisotropy parameter quantifying the degree of angular dependence, and $\theta_{\Sigma^+}$ represents the polar angle of the $\Sigma^+$ measured relative to the positron beam direction in the c.m.\ frame.
In this analysis, $\alpha_\theta$ is set to be  $-0.508$~\cite{BESIII:2021wkr} in the vicinity of the $J/\psi$ peak, corresponding to c.m.\ energies between 3.08 and 3.12 GeV. Due to the lack of experimental measurements and limited statistics, $\alpha_\theta$ is set to be $\pm 1$ for signal MC simulation at other energy points.

\section{Event selection}
\label{ST-selection}
The signal process under study is $e^{+}e^{-} \to \Sigma^{+} \bar{\Sigma}^{-}$, where the decays $\Sigma^{+} \to p\pi^{0}$, $\bar{\Sigma}^{-} \to \bar{p} \pi^{0}$, and  $\pi^{0} \to \gamma\gamma$ are reconstructed.  To increase the detection efficiency, only one $\pi^{0}$ is required to be reconstructed, while explicit reconstruction of the other $\pi^{0}$ is not mandatory.
Therefore, the selected final state consists of $p \bar{p} \gamma \gamma$. Two charged tracks with opposite charges are identified as proton-antiproton candidates. Particle identification (PID), which combines measurements of the specific ionization  energy loss ($dE/dx$) and TOF to form likelihoods $\mathcal{L}(h)\; (h = p,K,\pi)$ for each hadron hypothesis, is implemented by requiring $\mathcal{L}(p)>\mathcal{L}(\pi)$ and $\mathcal{L}(p)>\mathcal{L}(K)$. 

A minimum of two isolated photon candidates with energy deposition greater than 25 MeV in the barrel region ($|\cos \theta|\textless0.8$) and 50 MeV in the end-cap regions ($0.86\textless|\cos \theta|\textless0.92$) are required for $\pi^{0}$ reconstruction. 
The $\pi^{0}$ candidate is selected by requiring the invariant mass of two photons ($M_{\gamma\gamma}$) to be within the range $(m_{\pi^0}-60)<M_{\gamma\gamma}<(m_{\pi^0}+40)$~MeV/$c^2$, where $m_{\pi^0}$ is the nominal mass of the $\pi^0$ taken from the PDG~\cite{ParticleDataGroup:2024cfk}.
To suppress combinatorial background and improve mass resolution, a  kinematic fit is performed under the $e^{+}e^{-} \to p \bar{p} \pi^{0} \pi^{0}_{\rm {miss}}$ hypothesis by requiring total energy-momentum conservation and enforcing two constraints: $M_{\gamma\gamma}$ is constrained to the nominal $\pi^{0}$ mass, and $\pi^{0}_{\rm {miss}}$ is treated as a missing particle with fixed mass and unknown three-momentum.
The $\chi^{2}_{\rm {2C}}$ of the kinematic fit is required to be less than 20. The requirement is chosen by scanning the figure of merit
$S/\sqrt{S+B}$, where $S$ is the signal yield obtained from signal Monte Carlo (MC) simulation and $S+B$ is the total yield obtained from data. When an event contains more than two photons, the $\gamma \gamma$ combination with the lowest $\chi^{2}_{\rm{2C}}$ value is selected for $\pi^{0}$ reconstruction.

To determine whether the reconstructed $\pi^0$ is from the $\Sigma^{+}$ or $\bar{\Sigma}^{-}$ decay, the variables $\chi_{a}^{2}$ and $\chi_{a}^{b}$ are defined for the two possible combinations:
\begin{equation}
	\chi_{a}^{2} = (M_{p\pi^{0}}-m_{\Sigma^+})^2 + (M_{\bar{p}\pi^{0}_{\rm {miss}}}-m_{\bar{\Sigma}^{-}})^2,
	\label{chi1}
\end{equation}
\begin{equation}
	\chi_{b}^{2} = (M_{\bar{p}\pi^{0}}-m_{\bar{\Sigma}^{-}})^2 + (M_{p\pi^{0}_{\rm {miss}}}-m_{\Sigma^+})^2,
	\label{chi2}
\end{equation}
where $m_{\Sigma} = 1.189$ GeV/$c^{2}$ is the nominal $\Sigma^{+}/\bar{\Sigma}^{-}$ mass from the PDG~\cite{ParticleDataGroup:2024cfk}. The combination with the lower $\chi^{2} $ is retained. The invariant mass of the $\bar{\Sigma}^-$, reconstructed either via $\bar{p}\pi^{0}$ or $\bar{p}\pi^{0}_{\rm {miss}}$, is then required to be in the range $(1.17, 1.21)$~GeV/$c^{2}$.
 
\begin{figure*}[htbp]
	\centering
	\includegraphics[height=8cm]{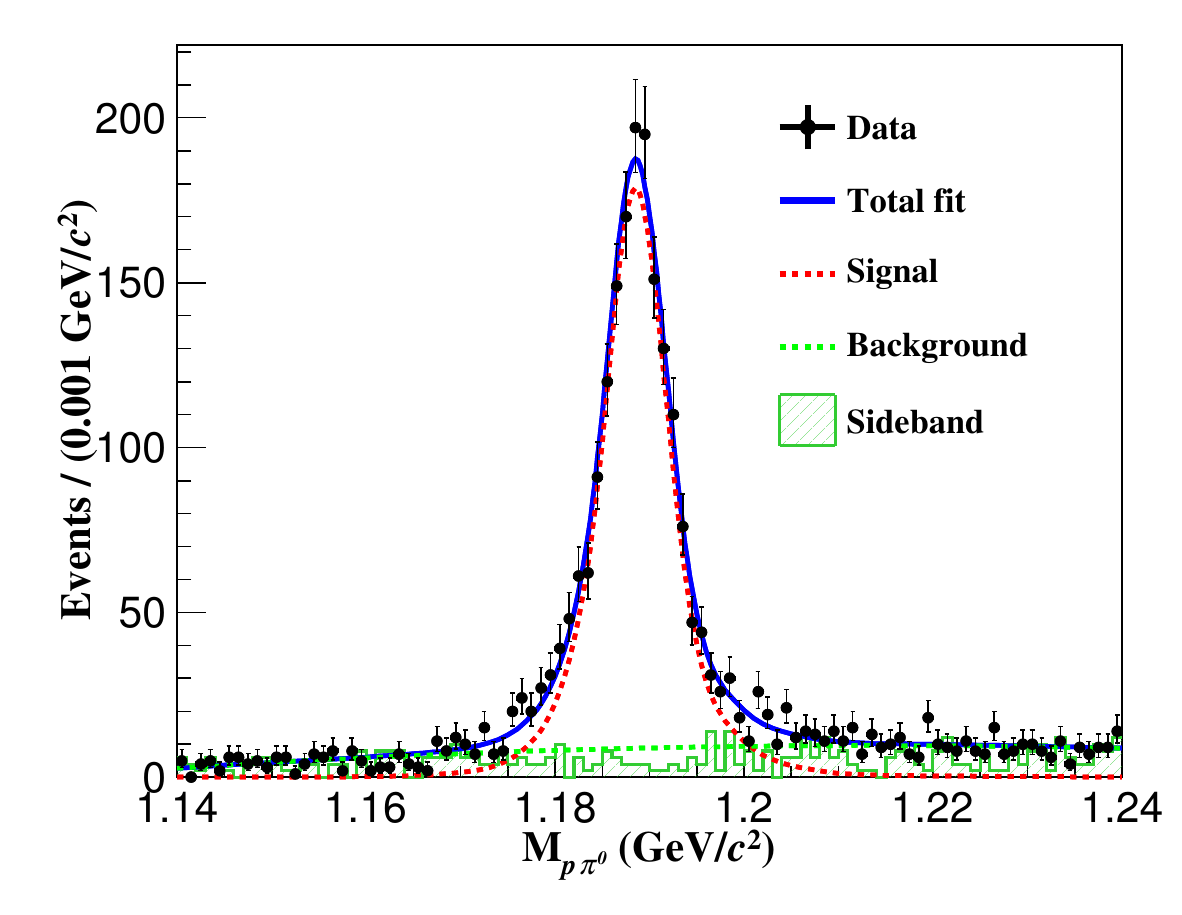}
	
	\caption{The fitted distribution of the invariant mass formed from either $p\pi^{0}$ or $p\pi^{0}_{\rm {miss}}$ at $\sqrt{s}$ = 3096.986~MeV showing a clear $\Sigma^+$ signal. The black dots with error bars are data. The blue solid line is the total fit result. The red dashed curve is the MC signal shape and the green dashed line is the background shape. The histograms filled with green diagonal lines represent the backgrounds estimated by the sidebands.}
	\label{fit_msigmap}
\end{figure*}

The signal is determined from the invariant mass of $\Sigma^{+}$ candidates, which is formed from either $p\pi^{0}$ or $p\pi^{0}_{\rm {miss}}$. Taking the data sample at $\sqrt{s}$ = 3096.986 MeV as an example, figure~\ref{fit_msigmap} presents the invariant mass distribution of $\Sigma^{+}$ candidates. The signal shape is modeled through the MC shape convolved with a Gaussian function, where the MC shape is constructed from signal MC candidates for which the matching angle between the reconstructed $\pi^0$ momentum and the MC-truth $\pi^0$ momentum is less than $10^{\circ}$. The background is described by a second-order Chebyshev polynomial. By performing an unbinned maximum likelihood fit, the number of signal events is obtained.

Both the inclusive MC sample and data sidebands are used to study potential background. The dominant backgrounds found in the inclusive MC sample include processes from $e^{+}e^{-}$ annihilation events with $\Delta$, $\Sigma$ and $\Lambda$ baryons. The sum of all backgrounds from the inclusive MC sample shows a smooth behavior, with no peaking structures. 
The data sideband regions are defined as $1.145\textless M_{\bar{\Sigma}^-} \textless 1.165$ GeV/$c^{2}$ and $1.215 \textless M_{\bar{\Sigma}^-}\textless 1.235$  GeV/$c^{2}$. As shown in figure~\ref{fit_msigmap}, the backgrounds from the sideband regions are smooth, and are consistent with those from  the inclusive MC sample.

\section{Determination of the observed cross section}

\subsection{Cross section calculation}

 The cross section for $e^{+}e^{-} \to \Sigma^{+} \bar{\Sigma}^{-}$ at a given c.m.\ energy is calculated via
\begin{equation}
	\sigma^{\rm obs}_{i}(e^{+}e^{-} \to \Sigma^{+} \bar{\Sigma}^{-})=\frac{N^{\rm sig}_{i}}{\mathcal{L}_i \cdot \varepsilon_i \cdot  \mathcal{B}_{\Sigma^{+} \to p \pi^0}^{2}},	
	\label{cs_sigma}
\end{equation}
where $N^{\rm sig}_{i}$ denotes the signal yield extracted from fitting the $M_{\Sigma^{+}}$ spectrum, $\mathcal{L}$ is the integrated luminosity, $\varepsilon$ is the detection efficiency given by the MC simulations, which will be illustrated later, and $\mathcal{B}_{\Sigma^{+} \to p \pi^0} = (51.47 \pm 0.30)\%$ represents the $\Sigma^+$ branching fraction taken from the PDG~\cite{ParticleDataGroup:2024cfk}. 
The specific values for these variables are summarized in table~\ref{table:dataset_and_obs_cs}. 

The detection efficiency at each c.m.\ energy is obtained based on the signal MC samples simulated with the {\sc ConExc}~\cite{Ping:2013jka} generator. 
Since the ISR distribution of simulated events depends on the input cross-section lineshape over a broad energy range, a composite lineshape is constructed as the generator input.
Specifically, the results obtained in this work (covering the region from 3.0000 to 3.1199 GeV) are merged with the lineshape from the threshold region reported in Ref.\cite{BESIII:2020uqk} to form a complete representation of the $e^{+}e^{-} \to \Sigma^{+} \bar{\Sigma}^{-}$ cross section.
To improve the estimation of the ISR effect, an iterative MC generation method is employed where the input cross-section lineshape is updated at each iteration. The iterations are repeated until convergence is achieved, which is defined as a relative difference of less than 0.5\% in the resulting cross section between the last two iterations.

\begin{table}[htbp]
	
		\caption{The integrated luminosity ($\mathcal{L}$), signal yield ($N^{\rm obs}$), detection efficiency ($\epsilon$), and observed cross section ($\sigma^{\rm obs}$) for the process $e^{+}e^{-} \to \Sigma^{+} \bar{\Sigma}^{-}$ at each c.m.\ energy ($\sqrt{s}$). The first and second uncertainties are statistical and 
		systematic for the observed cross sections, while statistical and systematic uncertainties are
		combined in quadrature for the c.m.\ energies~\cite{BESIII:2018wid} and luminosities~\cite{BESIII:2018wid, BESIII:2017lkp}. The data samples at the energy points marked with an asterisk ($^\ast$) were collected in 2012, while those marked with a dagger
		($^\dagger$) were collected in 2015.}
		\label{table:dataset_and_obs_cs}
		\resizebox{\textwidth}{!}{
			\begin{tabular}{lcccc} \hline
				$\sqrt{s}$ (MeV) & $\mathcal{L}$ (pb$^{-1}$) &
				$N^{\rm sig}$ & $\epsilon$ (\%) & $\sigma^{\rm obs}$ (pb) \\ \hline
				3000.00$\pm$0.20$^{\dagger}$ & 15.85$\pm$0.11 & $24.5\pm5.0$ & $36.0\pm0.1$ & $16.2\pm3.0\pm1.0$ \\
				3020.00$\pm$0.20$^{\dagger}$ & 17.32$\pm$0.12 & $27.9\pm5.8$ & $36.0\pm0.1$ & $16.9\pm3.5\pm1.1$ \\
				3049.66$\pm$0.03$^*$ & 14.92$\pm$0.16 & $27.6\pm5.7$ & $36.6\pm0.1$ & $19.1\pm3.9\pm1.1$ \\
				3058.71$\pm$0.03$^*$ & 15.06$\pm$0.16 & $19.0\pm4.8$ & $37.1\pm0.1$ & $12.8\pm3.2\pm0.7$ \\
				3079.65$\pm$0.03$^*$ & 17.51$\pm$0.20 & $29.0\pm6.8$ & $39.3\pm0.1$ & $15.9\pm3.7\pm0.9$ \\
				3080.00$\pm$0.20$^{\dagger}$ & 126.19$\pm$0.88 & $179.7\pm15.3$ & $39.8\pm0.1$ & $13.5\pm1.1\pm0.7$ \\
				3082.51$\pm$0.04$^*$ & 4.77$\pm$0.06 & $9.0\pm2.9$ & $40.0\pm0.1$ & $17.8\pm5.7\pm1.1$ \\
				3087.66$\pm$0.13 & 2.47$\pm$0.02 & $5.1\pm2.4$ & $43.1\pm0.1$ & $18.1\pm8.5\pm1.1$ \\
				3088.87$\pm$0.02$^*$ & 15.56$\pm$0.17 & $33.1\pm5.9$ & $43.4\pm0.1$ & $18.5\pm3.3\pm1.2$ \\
				3091.78$\pm$0.03$^*$ & 14.91$\pm$0.16 & $35.9\pm6.3$ & $45.7\pm0.1$ & $19.9\pm3.5\pm1.6$ \\
				3094.71$\pm$0.08$^*$ & 2.14$\pm$0.03 & $77.2\pm10.0$ & $54.8\pm0.1$ & $248.5\pm32.2\pm13.4$ \\
				3095.45$\pm$0.08$^*$ & 1.82$\pm$0.02 & $258.1\pm19.1$ & $55.0\pm0.1$ & $973.3\pm72.0\pm52.6$ \\
				3095.73$\pm$0.08 & 2.92$\pm$0.02 & $630.6\pm29.2$ & $56.2\pm0.1$ & $1450.5\pm67.2\pm91.4$ \\
				3095.84$\pm$0.08$^*$ & 2.14$\pm$0.03 & $649.6\pm29.3$ & $55.2\pm0.1$ & $2075.8\pm93.6\pm110.0$ \\
				3096.20$\pm$0.07 & 4.98$\pm$0.03 & $2102.2\pm54.0$ & $55.9\pm0.1$ & $2850.5\pm73.2\pm182.4$ \\
				3096.99$\pm$0.08 & 3.10$\pm$0.02 & $1767.3\pm50.0$ & $55.6\pm0.1$ & $3870.5\pm109.5\pm232.2$ \\
				3097.23$\pm$0.10 & 1.68$\pm$0.01 & $930.4\pm35.4$ & $56.0\pm0.1$ & $3733.1\pm142.0\pm242.6$ \\
				3097.23$\pm$0.08$^*$ & 2.07$\pm$0.03 & $1205.8\pm41.0$ & $54.8\pm0.1$ & $4012.5\pm136.4\pm212.7$ \\
				3097.65$\pm$0.08 & 4.66$\pm$0.03 & $2103.3\pm53.3$ & $56.1\pm0.1$ & $3037.0\pm77.0\pm188.3$ \\
				3098.36$\pm$0.08$^*$ & 2.20$\pm$0.03 & $535.3\pm26.7$ & $55.7\pm0.1$ & $1649.0\pm82.2\pm89.0$ \\
				3098.73$\pm$0.08 & 5.64$\pm$0.03 & $770.1\pm32.8$ & $56.2\pm0.1$ & $917.1\pm39.1\pm61.4$ \\
				3099.06$\pm$0.09$^*$ & 0.76$\pm$0.01 & $74.0\pm10.9$ & $55.2\pm0.1$ & $665.8\pm98.1\pm35.3$ \\
				3101.38$\pm$0.11$^*$ & 1.61$\pm$0.02 & $47.4\pm8.3$ & $53.6\pm0.1$ & $207.3\pm36.3\pm22.4$ \\
				3104.00$\pm$0.08 & 5.72$\pm$0.03 & $127.3\pm14.2$ & $52.8\pm0.1$ & $159.1\pm17.7\pm14.2$ \\
				3105.60$\pm$0.09$^*$ & 2.11$\pm$0.03 & $37.8\pm6.7$ & $51.4\pm0.1$ & $131.6\pm23.3\pm18.0$ \\
				3112.07$\pm$0.09$^*$ & 1.72$\pm$0.02 & $19.9\pm5.2$ & $47.8\pm0.1$ & $91.4\pm23.9\pm8.0$ \\
				3119.89$\pm$0.12$^*$ & 1.26$\pm$0.02 & $2.6\pm2.4$ & $40.8\pm0.1$ & $19.1\pm17.6\pm3.2$ \\ \hline
			\end{tabular}
		}
		
\end{table}

\subsection{Systematic uncertainties in the cross section}

The systematic uncertainties on the cross-section measurements stem from the components used in the calculation, as indicated by eq. (\ref{cs_sigma}). The uncertainties associated with $N_i^\text{sig}$ originate from the fit methodology and lineshape parameterization. The uncertainties on $\varepsilon_i$ arise from discrepancies between data and MC simulations regarding the event selection criteria, as well as the models used for MC generation, such as angular distribution modeling. Finally, the uncertainties from $\mathcal{L}_i$ and $\mathcal{B}_{\Sigma^{+} \to p \pi^0}$ are also incorporated.

\begin{itemize}
\item \textbf{Luminosity}:
The integrated luminosity is measured using $e^+ e^- \to \gamma \gamma$ events resulting in a relative uncertainty of 1.0\%~\cite{BESIII:2017lkp}.

\item \textbf{Tracking and PID}:
The systematic uncertainties of the tracking and PID are both assigned as 1.0\% per track, determined using a control sample of $J/\psi \to p\bar{p}\pi^{+}\pi^{-}$ events~\cite{BESIII:2012koo}.

\item \textbf{Photon reconstruction}: 
The systematic uncertainty associated with the photon reconstruction is assigned as 1.0\% per photon using a control sample of $J/\psi \to \pi^{+}\pi^{-}\pi^0$ events~\cite{BESIII:2018rdg}. For the final state containing two photons, an uncertainty of 2.0\% is assigned.

\item \textbf{Quoted branching fraction}: 
The branching fraction is quoted from the PDG~\cite{ParticleDataGroup:2024cfk}: $\Sigma^+ \to p \pi^0$ is $(51.47\pm0.30) \%$, with a relative uncertainty of 0.6\%. A total systematic uncertainty of 1.2\% is assigned. 

\item \textbf{$\Sigma^+$ angular distribution}: 
For $\sqrt{s} = 3.000-3.0796$ GeV, where the $\Sigma^+$ decay angular parameter $\alpha_{\theta}$ is poorly constrained, two extreme scenarios ($\alpha_{\theta} =\pm1 $) are tested in MC simulations. The difference in detection efficiency between these two extreme cases is considered as the maximum possible variation. Assuming a uniform probability distribution between these extremes,  the efficiency difference is divided by $\sqrt{12}$~\cite{angular} and assigned as the uncertainty. For $\sqrt{s} = 3.080 - 3.1199$ GeV, an alternative value of $\alpha_{\theta} =-0.485 $ is obtained from an angular analysis using data samples in the $J/\psi$ peak region.  This value is tested with MC simulations and the difference in the detection efficiency with respect to the nominal value ($\alpha_{\theta} =-0.508$)  is regarded as the systematic uncertainty at each energy point, which is shown in table~\ref{table:sys}.   

\item \textbf{$\bar{\Sigma}^{-}$ mass window}: 
The uncertainty associated with the $\bar{\Sigma}^{-}$ mass window is studied with a control sample of $J/\psi \to \Sigma^{+} \bar{\Sigma}^{-}$~\cite{BESIII:2021wkr}. The difference of efficiencies between data and MC simulation, 1.5\%, is assigned as the systematic uncertainty.

\item \textbf{Kinematic fit}: 
The uncertainty associated with the kinematic fit is studied with a control sample of $J/\psi \to \Sigma^{+} \bar{\Sigma}^{-}$~\cite{BESIII:2021wkr}. The difference of efficiencies between data and MC simulation, 2.9\%, is assigned as the systematic uncertainty.

\item \textbf{Fit procedure}: 
(1) $\Sigma^+\; fit\; range$: The invariant mass fit range is narrowed from (1.14,1.24)~GeV/$c^2$ to (1.146,1.234)~GeV/$c^2$. The shift in extracted signal yields across all c.m.\ energy points defines the systematic uncertainties. (2) $Signal\; shape$: The sensitivity to the MC match angle of the $\pi^0$, which is used to estimate the $\Sigma^+$ signal shape,  is evaluated by varying the selection window by  $\pm 5^{\circ}$. The maximum difference in signal yield  is taken as the systematic uncertainty. (3) $background\; shape$: The nominal second-order polynomial background model is replaced with a third-order polynomial. The difference in signal yield is assigned as the systematic uncertainty at each energy point, which is shown in table~\ref{table:sys}. The total systematic uncertainty in the fitting procedure is taken as the quadratic sum of the three components.

\end{itemize}

\begin{table*}[htp]
	
	\begin{center}
		\caption{Relative systematic uncertainties (in \%) from each component. Correlated systematic uncertainties are marked with *, and the others are independent. }
		\label{table:sys}
		\resizebox{\textwidth}{!}{
			\begin{tabular}{cccccccccccc} \hline
				$\sqrt{s}$ (MeV)& Lum.$^*$  &  PID$^*$ &Trk$^*$  & $\gamma$ Rec.$^*$ & Kine. fit$^*$ & $M_{\bar{\Sigma}^-}^*$ &Match $\pi^0$ &Ang. & Br$^*$ & Fit &Total  \\  \hline
				3000.00$\pm$0.20 &1.0 &2.0 &2.0 &2.0 &2.9 &1.5 &1.9 &2.3 &1.2 &1.7 & 6.1 \\  \hline
				3020.00$\pm$0.20 &1.0 &2.0 &2.0 &2.0 &2.9 &1.5 &1.0 &2.2 &1.2 &3.3 & 6.5 \\  \hline
				3049.66$\pm$0.03 &1.0 &2.0 &2.0 &2.0 &2.9 &1.5 &1.9 &1.5 &1.2 &1.8 & 5.8 \\  \hline
				3058.71$\pm$0.03 &1.0 &2.0 &2.0 &2.0 &2.9 &1.5 &0.9 &1.5 &1.2 &1.7 & 5.6 \\  \hline
				3079.65$\pm$0.03 &1.0 &2.0 &2.0 &2.0 &2.9 &1.5 &2.4 &0.9 &1.2 &1.8 & 5.9 \\  \hline
				3080.00$\pm$0.20 &1.0 &2.0 &2.0 &2.0 &2.9 &1.5 &0.5 &0.7 &1.2 &1.7 & 5.4 \\  \hline
				3082.51$\pm$0.04 &1.0 &2.0 &2.0 &2.0 &2.9 &1.5 &2.2 &1.3 &1.2 &1.7 & 5.9 \\  \hline
				3087.59$\pm$0.13 &1.0 &2.0 &2.0 &2.0 &2.9 &1.5 &2.4 &0.6 &1.2 &2.8 & 6.2 \\  \hline
				3088.87$\pm$0.02 &1.0 &2.0 &2.0 &2.0 &2.9 &1.5 &1.8 &1.9 &1.2 &3.5 & 6.6 \\  \hline
				3091.78$\pm$0.03 &1.0 &2.0 &2.0 &2.0 &2.9 &1.5 &1.1 &2.4 &1.2 &5.4 & 7.8 \\  \hline
				3094.71$\pm$0.08 &1.0 &2.0 &2.0 &2.0 &2.9 &1.5 &0.0 &0.7 &1.2 &1.8 & 5.4 \\  \hline
				3095.45$\pm$0.08 &1.0 &2.0 &2.0 &2.0 &2.9 &1.5 &0.3 &0.8 &1.2 &1.8 & 5.4 \\  \hline
				3095.73$\pm$0.08 &1.0 &2.0 &2.0 &2.0 &2.9 &1.5 &3.2 &1.0 &1.2 &1.7 & 6.3 \\  \hline
				3095.84$\pm$0.08 &1.0 &2.0 &2.0 &2.0 &2.9 &1.5 &0.3 &0.5 &1.2 &1.7 & 5.3 \\  \hline
				3096.20$\pm$0.07 &1.0 &2.0 &2.0 &2.0 &2.9 &1.5 &3.5 &0.7 &1.2 &1.7 & 6.4 \\  \hline
				3096.99$\pm$0.08 &1.0 &2.0 &2.0 &2.0 &2.9 &1.5 &2.9 &0.2 &1.2 &1.7 & 6.0 \\  \hline
				3097.23$\pm$0.10 &1.0 &2.0 &2.0 &2.0 &2.9 &1.5 &3.3 &1.7 &1.2 &1.7 & 6.5 \\  \hline
				3097.23$\pm$0.08 &1.0 &2.0 &2.0 &2.0 &2.9 &1.5 &0.4 &0.4 &1.2 &1.7 & 5.3 \\  \hline
				3097.65$\pm$0.08 &1.0 &2.0 &2.0 &2.0 &2.9 &1.5 &3.3 &0.1 &1.2 &1.7 & 6.2 \\  \hline
				3098.36$\pm$0.08 &1.0 &2.0 &2.0 &2.0 &2.9 &1.5 &0.5 &0.7 &1.2 &1.7 & 5.4 \\  \hline
				3098.73$\pm$0.08 &1.0 &2.0 &2.0 &2.0 &2.9 &1.5 &3.5 &2.3 &1.2 &1.7 & 6.7 \\  \hline
				3099.06$\pm$0.09 &1.0 &2.0 &2.0 &2.0 &2.9 &1.5 &0.3 &0.3 &1.2 &1.7 & 5.3 \\  \hline
				3101.38$\pm$0.11 &1.0 &2.0 &2.0 &2.0 &2.9 &1.5 &0.9 &0.7 &1.2 &9.5 & 10.8 \\  \hline
				3104.00$\pm$0.08 &1.0 &2.0 &2.0 &2.0 &2.9 &1.5 &0.8 &0.6 &1.2 &7.3 & 8.9 \\  \hline
				3105.60$\pm$0.09 &1.0 &2.0 &2.0 &2.0 &2.9 &1.5 &2.4 &0.7 &1.2 &12.5 & 13.7 \\  \hline
				3112.07$\pm$0.09 &1.0 &2.0 &2.0 &2.0 &2.9 &1.5 &0.8 &0.6 &1.2 &7.2 & 8.8 \\  \hline
				3119.89$\pm$0.12 &1.0 &2.0 &2.0 &2.0 &2.9 &1.5 &3.7 &2.4 &1.2 &15.5 & 16.9 \\  \hline
			\end{tabular}
		}
	\end{center}
\end{table*}

\section{Results}
\label{fit_results}
\subsection{Lineshape fit}

We perform a fit to the measured $e^{+}e^{-}\to\Sigma^{+}\bar{\Sigma}^{-}$
cross-section lineshape to determine the relative phase $\Phi_{3g,\gamma}$,
the branching fraction $\mathcal{B}(J/\psi\to\Sigma^{+}\bar{\Sigma}^{-})$,
and the continuum cross section. The observed line shape of the cross section is modified from the Born cross section due to effects of ISR and the beam energy spread, which is taken into account as described in the following.

Using the non-relativistic Breit-Wigner cross-section formula~\cite{Ceci:2013zta}, the Born cross section for $e^+ e^- \to \Sigma^{+} \bar{\Sigma}^{-}$ can be expressed as
  \begin{equation}
	\sigma_{0}=  \left| N_0 \frac{Se^{i\Phi} + E}{(M_{J/\psi}-W)-i(\Gamma_{J/\psi}/2)} - C \right|^2,
	\label{formula1}
\end{equation} 
where the normalization factor $N_0$ is $\sqrt{3\pi \Gamma_{J/\psi}^{2} B_{\rm{in}}[\frac{\hbar c}{W}]^2 \cdot 10^{10}}$, with $B_{\rm{in}}$ being the branching fraction of $e^+ e^- \to J/\psi$; $S$, $E$ and $C$ represent the strong, EM and continuum amplitudes, respectively; and $M_{J/\psi}$ and $\Gamma_{J/\psi}$ are the mass and width of the $J/\psi$ meson, respectively. 

In QED, the continuum cross section has an energy dependence given by $\sigma_{\rm cont}~\propto~W^{\rm -\lambda}$~\cite{1980fj}, where $W$ is the c.m.\ energy. This behavior can be parameterized with sufficient precision as:
\begin{equation}
	\sigma_{\rm cont}(W)=\sigma_{\rm cont.}(W_0)(\frac{W_0}{W})^{\lambda} = C^2,
\end{equation} 
where $\rm \lambda = 10$ is the energy exponent derived from QCD quark counting rules, which predicts a $W^{-5}$ scaling for the EM baryon amplitude due to form factor suppression. Here, $W_0 = 3.0 $ GeV serves as the reference energy, and  $\sigma_{\rm cont.}(W_0)$ denotes the cross-section at $W_0$.  The continuum amplitude $C$, expressed as a function of $W$ in MeV, is given by:
 \begin{equation}
	C=\sqrt{\sigma_{\rm cont.}(W_0)(\frac{W_0}{W})^{\rm \lambda}}.
\end{equation}  

To account for radiative effects in the c.m.\ energy, the experimentally observed cross section must be corrected for ISR. The radiatively corrected cross section becomes:
 \begin{equation}
	\sigma'(W)= \int_{0}^{1-(\frac{W_{\rm min}}{W})^2}dxF(W,x)\sigma^0(W\sqrt{1-x}),
	\label{ISR}
\end{equation} 
where $W_{\rm min}$ denotes the threshold of the $\Sigma^{+}\bar{\Sigma}^-$ system, $x=\frac{2E_{\gamma}}{W}$ parametrizes the fractional photon energy $E_{\gamma}$, and $F(W,x)$ represents the ISR photon emission probability. Adopting the structure function formalism from Ref.~\cite{Kuraev:1985hb}, the radiation kernel is approximated as: 
 \begin{equation}
	F(W,x)=x^{\beta-1}\beta \cdot (1+\delta)-\beta(1-\frac{x}{2})+\frac{1}{8}\beta^2 \left[ 4(2-x) \ln \frac{1}{x} - \frac{1+3(1-x)^2}{x} \ln (1-x)-6+x \right] ,
\end{equation} 
with radiative correction parameters: $\delta=\frac{3}{4}\beta+\frac{\alpha}{\pi}(\frac{\pi^2}{3}-\frac{1}{2})+\beta^2(\frac{9}{32}-\frac{\pi^2}{12})$ and $\beta =\frac{2\alpha}{\pi}(\ln\frac{W}{m^2_e}-1)$, 
where $\beta$ characterizes the effective bremsstrahlung coupling strength, and $\alpha$ denotes the fine structure constant.

To model experimental beam energy spread, the observed cross section in Eq.~\ref{ISR} must additionally be convolved with a Gaussian distribution characterized by the spread parameter $S_E$. The final cross-section becomes:

\begin{equation}
	\begin{split}
		\sigma_{\rm {exp}}(W)= \int_{W-nS_E}^{W+nS_E}\frac{1}{\sqrt{2\pi}S_E}e^{(\frac{-(W-W')^2}{2S^2_E})}\sigma'(W')dW'.
	\end{split}
	\label{fit_total}
\end{equation}

The parameters $\Phi_{3g,\gamma}$, $\mathcal{B}_{e^+ e^- \to \Sigma^+\bar{\Sigma}^-}$ and  $\sigma_{\rm cont.}(W_0)$ are determined through a covariance matrix minimization fit to $\sigma^{\rm obs}$ using the MINUIT package~\cite{Brun:1997pa}. The $\chi^{2}$ is defined as $\chi^2 = \Delta X^TM^{-1}\Delta X$, where $\Delta X$ represents the difference between data and theoretical predictions. The covariance matrix($M$) is constructed via: 
\begin{equation}
	\label{matrix}
M_{ij} = \sum_{k} (x_i \cdot \epsilon_{ik}) (x_j \cdot \epsilon_{jk}), 
\end{equation} 
where $x_i$ is the observed cross section at the $i$-th c.m.\ energy, and $\epsilon_{ik}$ is the relative systematic uncertainty from the $k$-th source for the $i$-th data point. 
Uncorrelated and correlated uncertainties are incorporated into the diagonal ($i=j$) and off-diagonal ($i \neq j$) elements of the covariance matrix, respectively.  The systematic uncertainties in the $e^+ e^- \to \Sigma^+ \bar{\Sigma}^-$ cross section measurements, categorized as correlated and uncorrelated components in table~\ref{table:sys}, are incorporated through a covariance matrix formalism for the $\chi^{2}$-Minimization function.

\subsection{Systematic uncertainties  for $\mathcal{B}_{e^+ e^- \to \Sigma^+\bar{\Sigma}^-}$  and $\Phi_{3g,\gamma}$}

Several sources of systematic uncertainties are considered in fitting the $e^+ e^- \to \Sigma^{+} \bar{\Sigma}^{-}$ cross section.  These include the energy spread, the ISR function, different lineshape descriptions, the continuum model and the effect of fixed parameters.

\begin{itemize}

\item \textbf{Energy spread}:
The uncertainty from the energy spread is evaluated by repeating the fit with $S_E$ treated as a free parameter. The systematic uncertainty of the measured branching fraction is extremely small and thus can be neglected, while the systematic uncertainty for $\Phi_{3g,\gamma}$ is determined to be 0.4\%.

\item \textbf{ISR function}:
The uncertainty due to the ISR function is estimated by considering several different definitions of the ISR photon emission probability. Since all approaches yield consistent results, no systematic uncertainty is assigned to the measured branching fraction and $\Phi_{3g,\gamma}$.

\item \textbf{Lineshape description}:
An alternative analytic fitting approach, which parametrizes the lineshape directly, as described in Ref.~\cite{BESIII:2018wid}, is employed to cross-check the nominal results. This provides a different mathematical treatment of the interference and resolution effects. Neither the phase nor branching fraction changes, and thus this systematic uncertainty is neglected.

\item \textbf{Continuum model}:
The uncertainty from the continuum parametrization is evaluated by floating the energy-dependent exponent $\lambda$. The systematic uncertainty is estimated to be  0.9\% for the measured branching fraction, and 0.4\% for $\Phi_{3g,\gamma}$.

\item \textbf{Fixed parameters}:
Input parameter uncertainties are propagated by varying $M_{J/\psi}$, $\Gamma_{J/\psi}$, $\mathcal{B}_{\mu\mu}$ and $\mathcal{B}_{ee}$ within their $\pm 1$ experimental uncertainties. Contributions from $M_{J/\psi}, \mathcal{B}_{\mu\mu}$ and $\mathcal{B}_{ee}$ are found to be negligible. The uncertainty due to $\Gamma_{J/\psi}$ is estimated to be 2.6\% for the measured branching fraction, and 0.4\%  for $\Phi_{3g,\gamma}$.
\end{itemize}

\begin{figure*}[htbp]
	
	\centering

	\begin{minipage}[b]{0.49\textwidth}
		\centering
		\includegraphics[width=\textwidth,height=5.5cm]{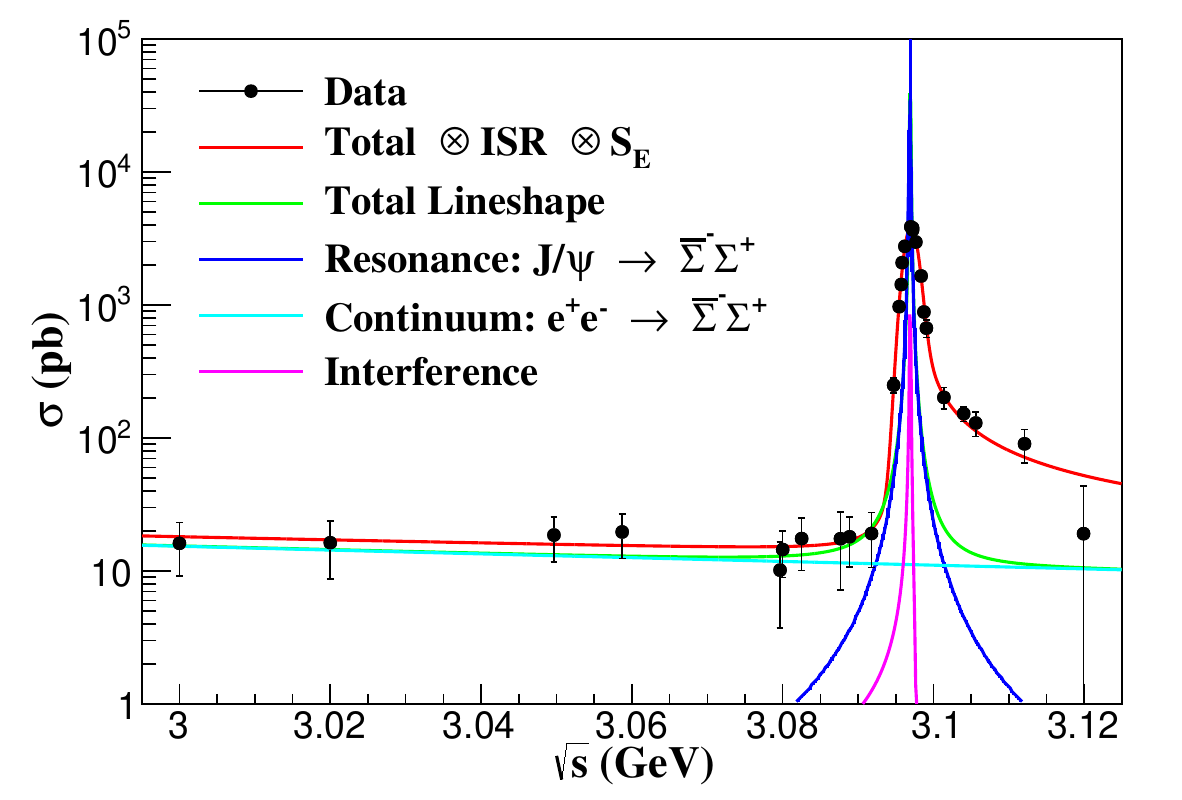}
		\centerline{(a)} 
	\end{minipage}
	\hfill
	\begin{minipage}[b]{0.49\textwidth}
		\centering
		\includegraphics[width=\textwidth,height=5.5cm]{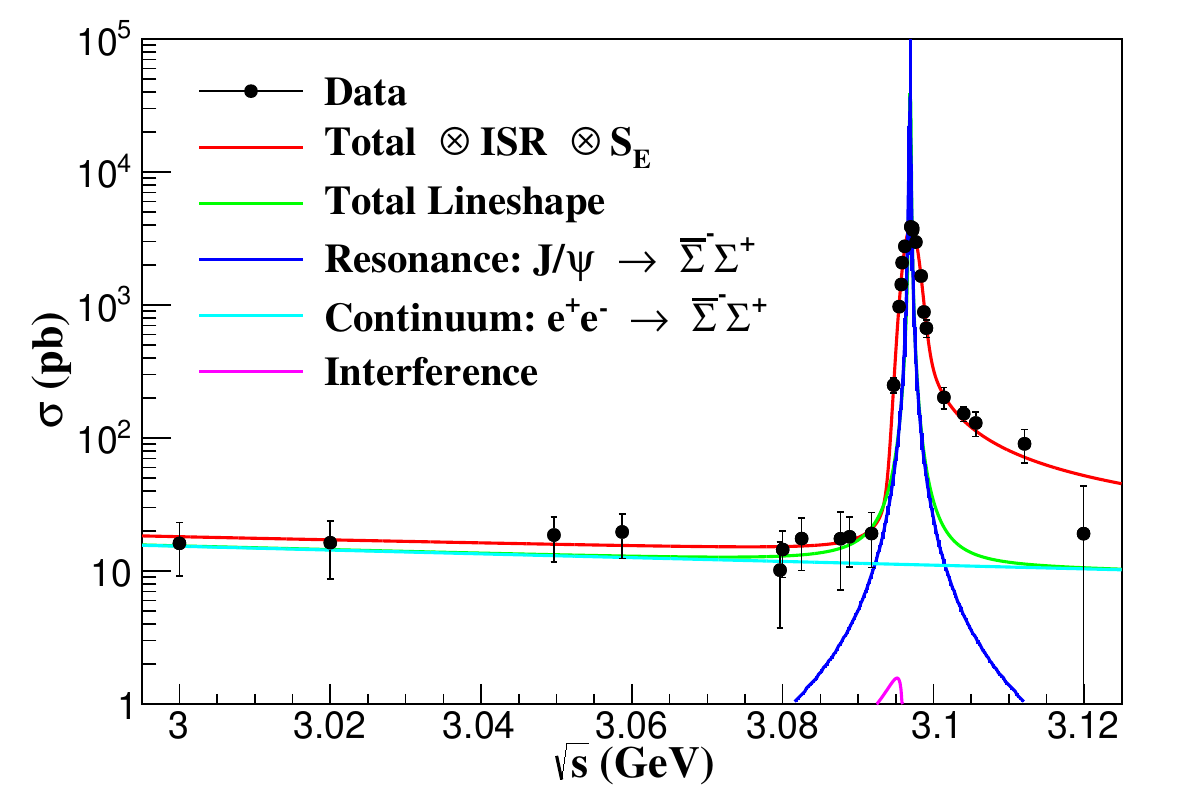}
		\centerline{(b)} 
	\end{minipage}
	
	\vspace{0.2cm}

	\begin{minipage}[b]{0.49\textwidth}
		\centering
		\includegraphics[width=\textwidth,height=5.5cm]{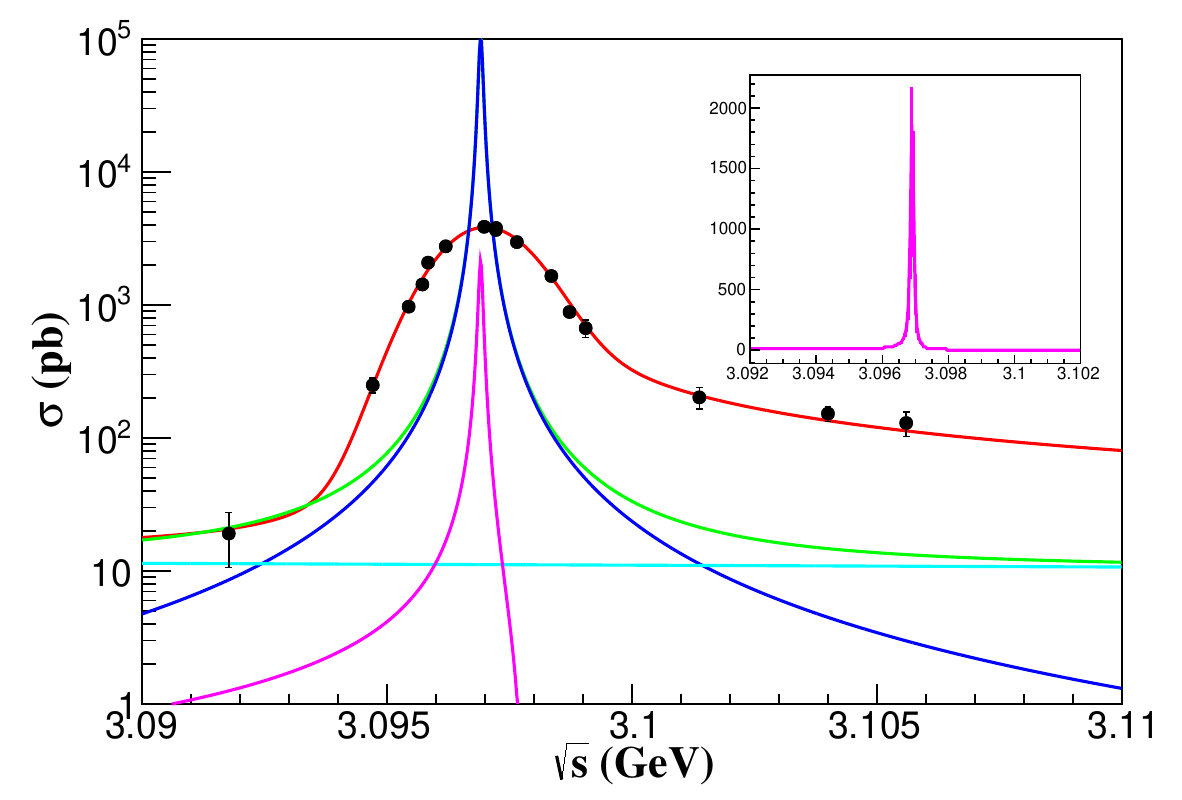}
		\centerline{(c)} 
	\end{minipage}
	\hfill
	\begin{minipage}[b]{0.49\textwidth}
		\centering
		\includegraphics[width=\textwidth,height=5.5cm]{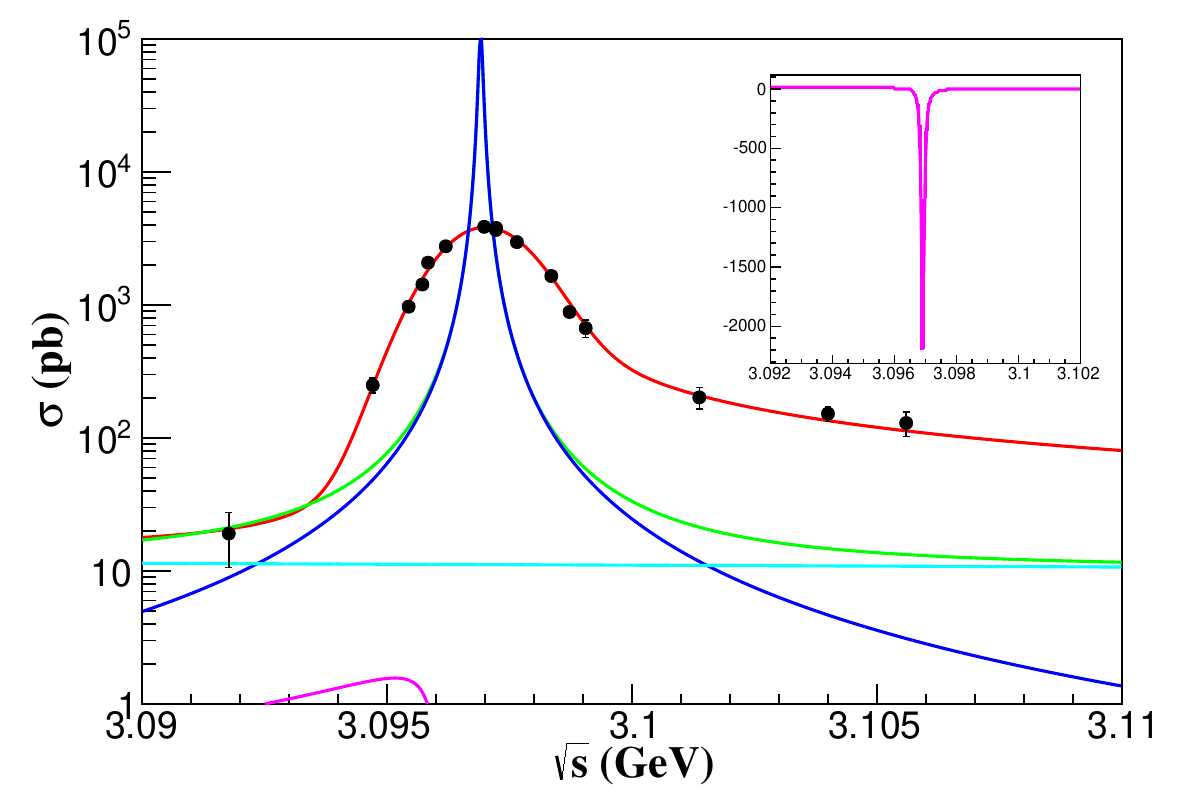}
		\centerline{(d)} 
	\end{minipage}
	
	\caption{Fits to the cross section lineshape for $e^+ e^- \to \Sigma^+\bar{\Sigma}^-$ under the positive phase  (a,c) and  negative phase (b,d) solutions. The black points with error bars are data. The red curve denotes the overall fit result considering the effects of ISR and beam energy spread, while the green curve represents the lineshape without these two effects. The other curves show the individual contribution of each component without these two effects. Figures (c) and (d) zoom in around the $J/\psi$ resonance peak, providing a more detailed view of the interference patterns in the cross section lineshape.}
	\label{fit_cs}
\end{figure*}

\begin{figure*}[htbp]
	
	\centering

	\includegraphics[height=8cm]{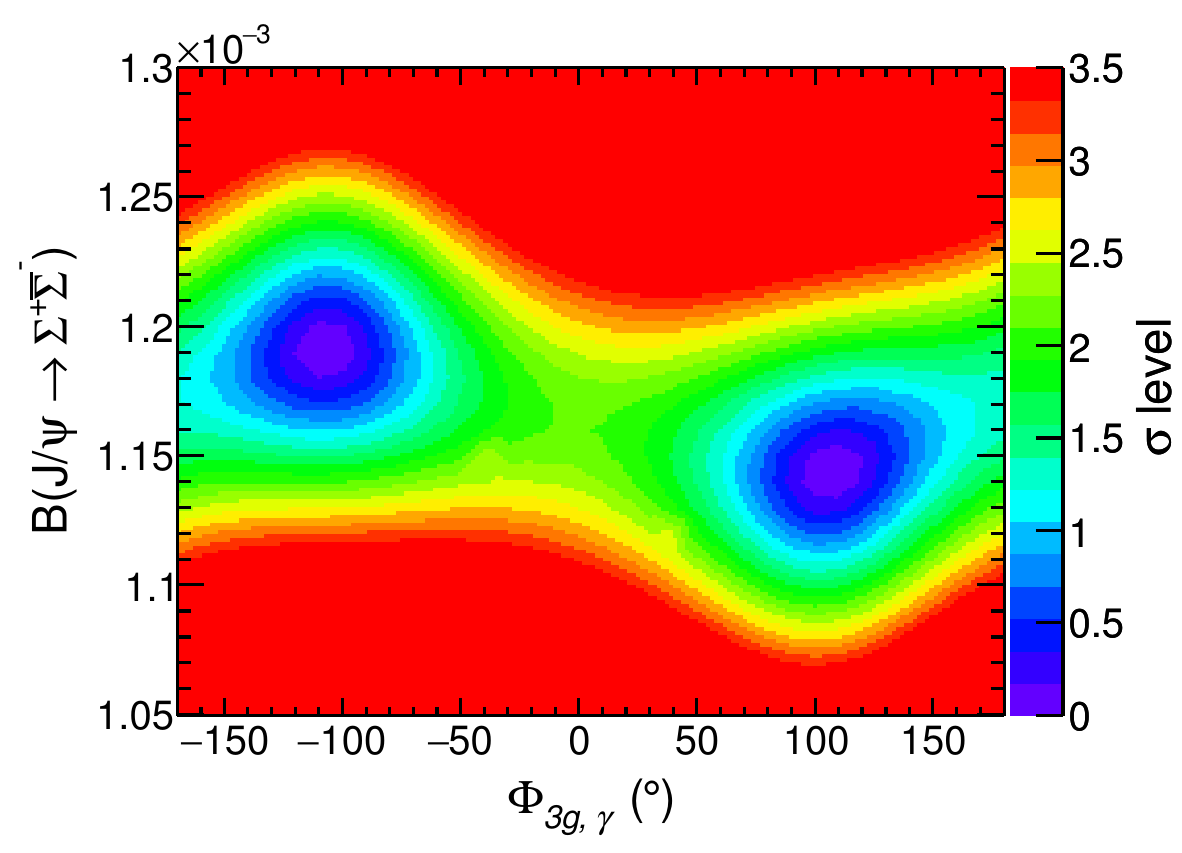}
	
	\caption{A 2D $\chi^{2}$-scan over a range of different values for $\Phi_{3g,\gamma}$ and $\mathcal{B}(J/\psi \to \Sigma^+ \bar{\Sigma}^-)$. The color scale indicates the statistical significance level in standard deviations. }
	\label{fit_chisq}
\end{figure*}
\subsection{Final results}

The final fit to the measured $e^+ e^- \to \Sigma^+\bar{\Sigma}^-$ observed cross-section includes the free parameters $\Phi_{3g,\gamma}$, $\mathcal{B}_{e^+ e^- \to \Sigma^+\bar{\Sigma}^-}$ and  $\sigma_{\rm cont.}(W_0)$.  It yields two solutions for the phase, $\Phi_{3g,\gamma}=(107.9 \pm 24.9)^\circ$ (the positive case) and $(-107.6 \pm 24.3)^\circ$ (the  negative case). The continuum cross-section at 3.000 GeV is determined to be  $\sigma_{\rm cont.}(W_0)= (15.4\pm3.1)$~pb for both solutions. All quoted uncertainties combine statistical and systematic components. 
Figure~\ref{fit_cs} displays the fitted lineshape, while table~\ref{table:total} summarizes the central values and uncertainties of all parameters determined by the fit. 
The resulting statistical significance contours for $\Phi_{3g,\gamma}$ versus $\mathcal{B}(J/\psi \to \Sigma^+ \bar{\Sigma}^-)$ are shown in figure~\ref{fit_chisq}. 
The scan reveals the two solutions as two dark blue regions, which demonstrate that the two solutions are well separated and easily distinguishable at the 1$\sigma$ confidence level.
The relative phase $\Phi_{3g,\gamma}$ is measured to be within the range [83.0$^\circ$, 133.5$^\circ$] and [$-132.8^\circ$, $-82.9^{\circ}$] within a 1$\sigma$ confidence interval.

\begin{table}[htp]
	
	\begin{center}
		\caption{Fit results of the cross section lineshape for  $e^+e^- \to \Sigma^+ \bar{\Sigma}^-$. Solution I corresponds to $\Phi_{3g,\gamma}>0^{\circ}$, and Solution II to $\Phi_{3g,\gamma}<0^{\circ}$, respectively. The uncertainties 
			arise from both the observed cross section and the phase fitting lineshape sources. ndf stands for the degrees of freedom used in calculating the $\chi^{2}$.}
		\label{table:total}
		\resizebox{\textwidth}{!}{
			\begin{tabular}{ccccc} \hline 
				\textbf{Solution}& $\Phi_{3g,\gamma}(^{\circ})$  &$\mathcal{B}(J/\psi \to \Sigma^+ \bar{\Sigma}^-)$ &$\sigma_{\rm cont.}$(3.000 GeV) pb & $\chi^{2}$/ndf   \\  \hline
				I &$\phantom{-}107.9 \pm 24.9$   & $(1.14\pm 0.04)\times 10^{-3}$&$15.4\pm3.1$& 17.4/24\\  
				II &$-107.6 \pm 24.3$  & $(1.19\pm 0.04)\times 10^{-3}$&$15.4\pm3.1$& 17.4/24\\  \hline

			\end{tabular}
		}
	\end{center}
\end{table}

\section{Summary}

Utilizing data collected at 27 c.m. energy points around the $J/\psi$ resonance,  we present the first direct measurement of the relative phase $\Phi_{3g,\gamma}$ between the strong and EM decay amplitudes for the process $e^+ e^- \to \Sigma^+\bar{\Sigma}^-$. 
Two solutions for $\Phi_{3g,\gamma}$ are obtained, $(107.9 \pm 24.9)^{ \circ }$ and $(-107.6 \pm 24.3)^{ \circ }$, corresponding to $\mathcal{B}_{J/\psi \to \Sigma^{+}\bar{\Sigma}^{-}} = (1.14 \pm 0.04)\times10^{-3}$ and $ (1.19 \pm 0.04)\times10^{-3}$, respectively. 
The measured value of $\Phi_{3g,\gamma}$ is consistent with the hypothesis of a universal orthogonal phase between the strong and EM amplitudes in $J/\psi$ decays~\cite{Wang:2002ef}, but shows significant deviation from the full interference predicted by pQCD models~\cite{Chernyak:1984bm, Brodsky:1987bb}. The obtained branching fraction are consistent with the current world average $(1.07 \pm 0.04)\times10^{-3}$~\cite{ParticleDataGroup:2024cfk}, within 3$\sigma$.  Furthermore, the measured $\sigma_{\rm cont.}(\rm 3.000)$ GeV agrees with the previous BESIII result~\cite{BESIII:2020uqk} within 1$\sigma$.
These results establish a benchmark for understanding the interplay between strong and EM interactions in charmonium decays.

\acknowledgments

The BESIII Collaboration thanks the staff of BEPCII (https://cstr.cn/31109.02.BEPC) and the IHEP computing center for their strong support. This work is supported in part by National Key R\&D Program of China under Contracts Nos. 2023YFA1606000, 2023YFA1606704, 2025YFA1613900, 2023YFA1609400; Natural Science Foundation of China (NSFC) under Contracts Nos. 11635010, 11935015, 11935016, 11935018, 12025502,  12035009, 12035013, 12061131003, 12105100, 12105276, 12192260, 12192261, 12192262, 12192263, 12192264, 12192265, 12221005, 12225509, 12235017, 12342502, 12361141819, 12535005, 12625507;  The Natural Science Foundation of Hunan Province under Contract No. 2024JJ2044; Beijing Natural Science Foundation under Contracts No. JQ22002; The Fundamental Research Funds for the Central Universities (2026MS077); Joint Large-Scale Scientific Facility Funds of the NSFC and CAS under Contracts Nos. U2032111; The Chinese Academy of Sciences (CAS) Large-Scale Scientific Facility Program; The Strategic Priority Research Program of Chinese Academy of Sciences under Contract No. XDA0480600; CAS under Contract No. YSBR-101; 100 Talents Program of CAS; The Institute of Nuclear and Particle Physics (INPAC) and Shanghai Key Laboratory for Particle Physics and Cosmology; ERC under Contract No. 758462; German Research Foundation DFG under Contract No. FOR5327; Istituto Nazionale di Fisica Nucleare, Italy; Knut and Alice Wallenberg Foundation under Contracts Nos. 2021.0174, 2021.0299, 2023.0315; Ministry of Development of Turkey under Contract No. DPT2006K-120470; National Research Foundation of Korea under Contract No. NRF-2022R1A2C1092335; National Science and Technology fund of Mongolia; Polish National Science Centre under Contract No. 2024/53/B/ST2/00975; STFC (United Kingdom); Swedish Research Council under Contract No. 2019.04595; U. S. Department of Energy under Contract No. DE-FG02-05ER41374.\\

\bibliographystyle{JHEP}
\bibliography{references}

@article{SLAC-SP-017:1974ind,
	author = "Augustin, J. E. and others",
	collaboration = "SLAC-SP-017",
	title = "{Discovery of a Narrow Resonance in $e^+ e^-$ Annihilation}",
	reportNumber = "SLAC-PUB-1504, LBL-3391",
	doi = "10.1103/PhysRevLett.33.1406",
	journal = "Phys. Rev. Lett.",
	volume = "33",
	pages = "1406--1408",
	year = "1974"
}

@article{Freund:1975pn,
	author = "Freund, Peter G. O. and Nambu, Yoichiro",
	title = "{Dynamics in the Zweig Iizuka Rule and a New Vector Meson Below 2 $GeV/c^2$}",
	reportNumber = "EFI 75/3",
	doi = "10.1103/PhysRevLett.34.1645",
	journal = "Phys. Rev. Lett.",
	volume = "34",
	pages = "1645",
	year = "1975"
}

@article{MARK-III:1988crp,
	author = "Coffman, D. and others",
	collaboration = "MARK-III",
	title = "{Measurements of $J/\psi$ Decays Into a Vector and a Pseudoscalar Meson}",
	reportNumber = "SLAC-PUB-4424",
	doi = "10.1103/PhysRevD.38.2695",
	journal = "Phys. Rev. D",
	volume = "38",
	pages = "2695",
	year = "1988"
}

@article{Wang:2002ef,
	author = "Wang, P. and Yuan, C. Z. and Mo, X. H. and Zhang, D. H.",
	title = "{The Interference between continuum and resonance in $e^+ e^- \to c \bar{c}$ experiment}",
	eprint = "hep-ph/0212139",
	archivePrefix = "arXiv",
	month = "12",
	year = "2002"
}

@article{Czyz:2009vj,
	author = "Czyz, Henryk and Kuhn, Johann H.",
	title = "{Strong and Electromagnetic $J/\psi$ and $\psi(2S)$ Decays into Pion and Kaon Pairs}",
	reportNumber = "TTP09-11",
	doi = "10.1103/PhysRevD.80.034035",
	journal = "Phys. Rev. D",
	volume = "80",
	pages = "034035",
	year = "2009"
}

@article{Zhou:2017yod,
	author = "Zhou, Xing Yu and Wang, Ya Di and Xia, Li Gang",
	title = "{Analytic forms for cross sections of di-lepton production from $e^+e^-$ collisions around the $J/\psi$ resonance}",
	doi = "10.1088/1674-1137/41/8/083001",
	journal = "Chin. Phys. C",
	volume = "41",
	number = "8",
	pages = "083001",
	year = "2017"
}

@article{Guo:2022gkg,
	author = "Guo, Y. P. and Yuan, C. Z.",
	title = "{Impact of the interference between the resonance and continuum amplitudes on vector quarkonia decay branching fraction measurements}",
	doi = "10.1103/PhysRevD.105.114001",
	journal = "Phys. Rev. D",
	volume = "105",
	number = "11",
	pages = "114001",
	year = "2022"
}

@article{BESIII:2018wid,
	author = "Ablikim, M. and others",
	collaboration = "BESIII",
	title = "{Measurement of the phase between strong and electromagnetic amplitudes of $J/\psi$ decays}",
	doi = "10.1016/j.physletb.2019.03.001",
	journal = "Phys. Lett. B",
	volume = "791",
	pages = "375--384",
	year = "2019"
}

@article{Brodsky:1987bb,
	author = "Brodsky, Stanley J. and Lepage, G. Peter and Tuan, S. F.",
	title = "{Exclusive Charmonium Decays: The $J/\psi (\psi^\prime$) $\to \rho \pi$, $K^* \bar{K}$ Puzzle}",
	reportNumber = "SLAC-PUB-4276, UH-511-618-87",
	doi = "10.1103/PhysRevLett.59.621",
	journal = "Phys. Rev. Lett.",
	volume = "59",
	pages = "621",
	year = "1987"
}

@article{LopezCastro:1994xw,
	author = "L\'opez Castro, G. and Lucio M., J. L. and Pestieau, J.",
	title = "{Tests of flavor symmetry in $J/\psi$ decays}",
	doi = "10.1063/1.48783",
	journal = "AIP Conf. Proc.",
	volume = "342",
	pages = "441--448",
	year = "1995"
}

@article{Baldini:1998en,
	author = "Baldini, R and others",
	title = "{Measurement of $J / \psi \to N \bar{N}$ branching ratios and estimate of the phase of the strong decay amplitude}",
	doi = "10.1016/S0370-2693(98)01358-6",
	journal = "Phys. Lett. B",
	volume = "444",
	pages = "111--118",
	year = "1998"
}

@article{Zhu:2015bha,
	author = "Zhu, Kai and Mo, Xiao Hu and Yuan, Chang Zheng",
	title = "{Determination of the relative phase in $\psi'$ and $J/\psi$ decays into baryon and antibaryon}",
	doi = "10.1142/S0217751X15501481",
	journal = "Int. J. Mod. Phys. A",
	volume = "30",
	number = "25",
	pages = "1550148",
	year = "2015"
}

@article{DM2:1988bfq,
	author = "Jousset, J. and others",
	collaboration = "DM2",
	title = "{The $J/\psi \to$ Vector + Pseudoscalar Decays and the $\eta$, $\eta^\prime$ Quark Content}",
	reportNumber = "LAL-88-25",
	doi = "10.1103/PhysRevD.41.1389",
	journal = "Phys. Rev. D",
	volume = "41",
	pages = "1389",
	year = "1990"
}

@article{Suzuki:1999nb,
	author = "Suzuki, M.",
	title = "{A large final state interaction in the $0^-0^-$ decays of $J/\psi$}", 
	reportNumber = "LBNL-42731, UCB-PTH-99-01, LBL-42731",
	doi = "10.1103/PhysRevD.60.051501",
	journal = "Phys. Rev. D",
	volume = "60",
	pages = "051501",
	year = "1999"
}

@article{Kopke:1988cs,
	author = "Kopke, L. and Wermes, N.",
	title = "{$J/\psi$ Decays}",
	reportNumber = "CERN-EP-88-93",
	doi = "10.1016/0370-1573(89)90074-4",
	journal = "Phys. Rept.",
	volume = "174",
	pages = "67",
	year = "1989"
}

@article{Chernyak:1984bm,
	author = "Chernyak, V. L. and Zhitnitsky, I. R.",
	title = "{Nucleon Wave Function and Nucleon Form-Factors in QCD}",
	doi = "10.1016/0550-3213(84)90114-7",
	journal = "Nucl. Phys. B",
	volume = "246",
	pages = "52--74",
	year = "1984"
}

@article{Li:2017jpg,
	author = "Li, Xin and others",
	title = "{Study of MRPC technology for BESIII endcap-TOF upgrade}",
	doi = "10.1007/s41605-017-0014-2",
	journal = "Radiat. Detect. Technol. Methods",
	volume = "1",
	pages = "13",
	year = "2017"
}

@article{Guo:2017sjt,
	author = "Guo, Ying Xiao and others",
	title = "{The study of time calibration for upgraded end cap TOF of BESIII}",
	doi = "10.1007/s41605-017-0012-4",
	journal = "Radiat. Detect. Technol. Methods",
	volume = "1",
	pages = "15",
	year = "2017"
}

@article{Cao:2020ibk,
	author = "Cao, P. and others",
	title = "{Design and construction of the new BESIII endcap Time-of-Flight system with MRPC Technology}",
	doi = "10.1016/j.nima.2019.163053",
	journal = "Nucl. Instrum. Meth. A",
	volume = "953",
	pages = "163053",
	year = "2020"
}

@article{Yang:2014vra,
	author = "Yang, Rui Ling and Ping, Rong Gang and Chen, Hong",
	title = "{Tuning and Validation of the Lundcharm Model with $J/\psi$ Decays}",
	doi = "10.1088/0256-307X/31/6/061301",
	journal = "Chin. Phys. Lett.",
	volume = "31",
	pages = "061301",
	year = "2014"
}

@article{Suzuki:2000yq,
	author = "Suzuki, Mahiko",
	title = "{Anomalies in $\psi(2S)$ decay and the $\rho\pi$ puzzle}",
	reportNumber = "LBNL-46203, LBL-46203, UCB-PTH-00-20",
	doi = "10.1103/PhysRevD.63.054021",
	journal = "Phys. Rev. D",
	volume = "63",
	pages = "054021",
	year = "2001"
}

@article{CLEO:2006udi,
	author = "Dobbs, S. and others",
	collaboration = "CLEO",
	title = "{Measurement of interference between electromagnetic and strong amplitudes in $\psi(2S)$ decays to two pseudoscalar mesons}",
	reportNumber = "CLNS-05-1947, CLEO-05-33",
	doi = "10.1103/PhysRevD.74.011105",
	journal = "Phys. Rev. D",
	volume = "74",
	pages = "011105",
	year = "2006"
}

@article{Brun:1997pa,
	author = "Brun, R. and Rademakers, F.",
	editor = "Werlen, M. and Perret-Gallix, D.",
	title = "{ROOT: An object oriented data analysis framework}",
	doi = "10.1016/S0168-9002(97)00048-X",
	journal = "Nucl. Instrum. Meth. A",
	volume = "389",
	pages = "81--86",
	year = "1997"
}

@article{Chen:1998ma,
	author = "Chen, Yu Qi and Braaten, Eric",
	title = "{An Explanation for the $\rho\pi$ puzzle of $J /\psi$ and $\psi'$ decays}",
	reportNumber = "OHSTPY-TH-98-001",
	doi = "10.1103/PhysRevLett.80.5060",
	journal = "Phys. Rev. Lett.",
	volume = "80",
	pages = "5060--5063",
	year = "1998"
}

@article{BESIII:2009fln,
	author = "Ablikim, M. and others",
	collaboration = "BESIII",
	title = "{Design and Construction of the BESIII Detector}",
	primaryClass = "physics.ins-det",
	doi = "10.1016/j.nima.2009.12.050",
	journal = "Nucl. Instrum. Meth. A",
	volume = "614",
	pages = "345--399",
	year = "2010"
}

@inproceedings{Yu:2016cof,
	author       = {Yu, Chenghui and others},
	title        = {{BEPCII} Performance and Beam Dynamics Studies on Luminosity},
	booktitle    = {\href{https://jacow.org/ipac2016/papers/tuya01.pdf}{7th International Particle Accelerator Conference \rm{ \textbf{TUYA01} (2016)}}},

}

@article{Huang:2022wuo,
	author = "Huang, Kai Xuan and Li, Zhi Jun and Qian, Zhen and Zhu, Jiang and Li, Hao Yuan and Zhang, Yu Mei and Sun, Sheng Sen and You, Zheng Yun",
	title = "{Method for detector description transformation to Unity and application in BESIII}",
	primaryClass = "physics.ins-det",
	doi = "10.1007/s41365-022-01133-8",
	journal = "Nucl. Sci. Tech.",
	volume = "33",
	number = "11",
	pages = "142",
	year = "2022"
}

@article{Abakumova:2011rp,
	author = "Abakumova, E. V. and others",
	title = "{The Beam energy measurement system for the Beijing electron-positron collider}",
	primaryClass = "physics.acc-ph",
	doi = "10.1016/j.nima.2011.08.050",
	journal = "Nucl. Instrum. Meth. A",
	volume = "659",
	pages = "21--29",
	year = "2011"
}

@article{ParticleDataGroup:2024cfk,
	author = "Navas, S. and others",
	collaboration = "Particle Data Group",
	title = "{Review of particle physics}",
	doi = "10.1103/PhysRevD.110.030001",
	journal = "Phys. Rev. D",
	volume = "110",
	number = "3",
	pages = "030001",
	year = "2024"
}

@article{BESIII:2017lkp,
	author = "Ablikim, M. and others",
	collaboration = "BESIII",
	title = "{Luminosity measurements for the R scan experiment at BESIII}",
	doi = "10.1088/1674-1137/41/6/063001",
	journal = "Chin. Phys. C",
	volume = "41",
	number = "6",
	pages = "063001",
	year = "2017"
}

@article{GEANT4:2002zbu,
	author = "Agostinelli, S. and others",
	collaboration = "GEANT4",
	title = "{GEANT4--a simulation toolkit}",
	reportNumber = "SLAC-PUB-9350, FERMILAB-PUB-03-339, CERN-IT-2002-003",
	doi = "10.1016/S0168-9002(03)01368-8",
	journal = "Nucl. Instrum. Meth. A",
	volume = "506",
	pages = "250--303",
	year = "2003"
}

@article{Ping:2013jka,
	author = "Ping, Rong Gang",
	title = "{An exclusive event generator for $e^+ e^-$ scan experiments}",
	
	doi = "10.1088/1674-1137/38/8/083001",
	journal = "Chin. Phys. C",
	volume = "38",
	pages = "083001",
	year = "2014"
}

@article{Ping:2008zz,
	author = "Ping, Rong Gang",
	title = "{Event generators at BESIII}",
	doi = "10.1088/1674-1137/32/8/001",
	journal = "Chin. Phys. C",
	volume = "32",
	pages = "599",
	year = "2008"
}

@article{Lange:2001uf,
	author = "Lange, D. J.",
	editor = "Erhan, S. and Schlein, P. and Rozen, Y.",
	title = "{The EvtGen particle decay simulation package}",
	doi = "10.1016/S0168-9002(01)00089-4",
	journal = "Nucl. Instrum. Meth. A",
	volume = "462",
	pages = "152--155",
	year = "2001"
}

@article{Richter-Was:1992hxq,
	author = "Richter Was, E.",
	title = "{QED bremsstrahlung in semileptonic B and leptonic tau decays}",
	reportNumber = "CERN-TH-6746-92",
	doi = "10.1016/0370-2693(93)90062-M",
	journal = "Phys. Lett. B",
	volume = "303",
	pages = "163--169",
	year = "1993"
}

@article{Nova:1990prh,
	author  = {Nova, S. and Olchevski, A. and Todorov, T.},
	title   = {{M}onte-{C}arlo Event Generator for Two-Photon Physics},
	journal = {\href{https://cds.cern.ch/record/184349}{DELPHI--90--35 PROG \rm {\textbf{152} (1990)}}},
}

@article{Balossini:2006wc,
	author = "Balossini, Giovanni and Carloni Calame, Carlo M. and Montagna, Guido and Nicrosini, Oreste and Piccinini, Fulvio",
	title = "{Matching perturbative and parton shower corrections to Bhabha process at flavour factories}",
	doi = "10.1016/j.nuclphysb.2006.09.022",
	journal = "Nucl. Phys. B",
	volume = "758",
	pages = "227--253",
	year = "2006"
}

@article{Balossini:2008xr,
	author = "Balossini, G. and Bignamini, C. and Calame, C. M. Carloni and Montagna, G. and Nicrosini, O. and Piccinini, F.",
	title = "{Photon pair production at flavour factories with per mille accuracy}",
	reportNumber = "FNT-T-2008-01, SHEP-08-05",
	doi = "10.1016/j.physletb.2008.04.007",
	journal = "Phys. Lett. B",
	volume = "663",
	pages = "209--213",
	year = "2008"
}

@article{Ceci:2013zta,
	author = "Ceci, S. and Korolija, M. and Zauner, B.",
	title = "{Model independent extraction of the pole and Breit-Wigner resonance parameters}",
	doi = "10.1103/PhysRevLett.111.112004",
	journal = "Phys. Rev. Lett.",
	volume = "111",
	pages = "112004",
	year = "2013"
}

@article{BESIII:2021wkr,
	author = "Ablikim, M. and others",
	collaboration = "BESIII",
	title = "{Measurement of branching fractions of~$J/\psi$ and $\psi(3686)$ decays to $\Sigma^{+}$ and $\overline{\Sigma}^-$}",
	doi = "10.1007/JHEP11(2021)226",
	journal = "JHEP",
	volume = "11",
	pages = "226",
	year = "2021"
}

@article{BESIII:2020uqk,
	author = "Ablikim, Medina and others",
	collaboration = "BESIII",
	title = "{Measurements of $\Sigma^+$ and $\Sigma^-$ time-like electromagnetic form factors for center-of-mass energies from 2.3864 to 3.0200 GeV}",
	doi = "10.1016/j.physletb.2021.136110",
	journal = "Phys. Lett. B",
	volume = "814",
	pages = "136110",
	year = "2021"
}

@article{1980fj,
	author = "Lepage, G. Peter and Brodsky, Stanley J.",
	title = "{Exclusive Processes in Perturbative Quantum Chromodynamics}",
	reportNumber = "SLAC-PUB-2478",
	doi = "10.1103/PhysRevD.22.2157",
	journal = "Phys. Rev. D",
	volume = "22",
	pages = "2157",
	year = "1980"
}

@article{Kuraev:1985hb,
	author  = {Kuraev, E. A. and Fadin, Victor S.},
	title   = {On Radiative Corrections to $e^+e^-$ Single Photon Annihilation at High Energy},
	journal = {\href{https://www.springer.com/journal/11988/volumes/41}{Sov.~J.~Nucl.~Phys. \rm {\textbf{41} (1985) 466--472}}}
}

@article{BESIII:2012koo,
	author = "Ablikim, M. and others",
	collaboration = "BESIII",
	title = "{Measurements of $\psi^\prime \to \bar{p} K^+ \Sigma^0$ and $\chi_{cj} \to \bar{p} K^+ \Lambda$}",
	doi = "10.1103/PhysRevD.87.012007",
	journal = "Phys. Rev. D",
	volume = "87",
	number = "1",
	pages = "012007",
	year = "2013"
}

@article{BESIII:2018rdg,
	author = "Ablikim, Medina and others",
	collaboration = "BESIII",
	title = "{Search for a strangeonium-like structure $Z_s$ decaying into $\phi \pi$ and a measurement of the cross section $e^+e^-\rightarrow\phi\pi\pi$}",
	doi = "10.1103/PhysRevD.99.011101",
	journal = "Phys. Rev. D",
	volume = "99",
	number = "1",
	pages = "011101",
	year = "2019"
}

@book{angular,
	author = "R. Wanke",
	title = "{Data Analysis in High Energy Physics}",
	publisher = "Wiley-VCH Verlag GmbH and Co. KGaA",
	place = {Singapore},
	year = "2013",
	pages = "263--280"
}

\clearpage
\appendix

M.~Ablikim$^{1}$\BESIIIorcid{0000-0002-3935-619X},
M.~N.~Achasov$^{4,c}$\BESIIIorcid{0000-0002-9400-8622},
P.~Adlarson$^{83}$\BESIIIorcid{0000-0001-6280-3851},
X.~C.~Ai$^{89}$\BESIIIorcid{0000-0003-3856-2415},
C.~S.~Akondi$^{31A,31B}$\BESIIIorcid{0000-0001-6303-5217},
R.~Aliberti$^{39}$\BESIIIorcid{0000-0003-3500-4012},
A.~Amoroso$^{82A,82C}$\BESIIIorcid{0000-0002-3095-8610},
Q.~An$^{78,65,\dagger}$,
Y.~H.~An$^{89}$\BESIIIorcid{0009-0008-3419-0849},
M.~S.~Anderson$^{39}$\BESIIIorcid{0009-0008-1550-2632}
% ...
\\
\vspace{0.2cm}
(BESIII Collaboration)\\
\vspace{0.2cm} {\it
$^{1}$ Institute of High Energy Physics, Beijing 100049, People's Republic of China\\
% ...
$^{4}$ Budker Institute of Nuclear Physics SB RAS (BINP), Novosibirsk 630090, Russia\\
% ...
$^{31}$ INFN Sezione di Ferrara, (A)INFN Sezione di Ferrara, I-44122, Ferrara, Italy; (B)University of Ferrara, I-44122, Ferrara, Italy\\
% ...
$^{39}$ Johannes Gutenberg University of Mainz, Johann-Joachim-Becher-Weg 45, D-55099 Mainz, Germany\\
% ...
$^{65}$ State Key Laboratory of Particle Detection and Electronics, Beijing 100049, Hefei 230026, People's Republic of China\\
% ...
$^{78}$ University of Science and Technology of China, Hefei 230026, People's Republic of China\\
% ...
$^{82}$ University of Turin and INFN, (A)University of Turin, I-10125, Turin, Italy; (B)University of Eastern Piedmont, I-15121, Alessandria, Italy; (C)INFN, I-10125, Turin, Italy\\
$^{83}$ Uppsala University, Box 516, SE-75120 Uppsala, Sweden\\
% ...
$^{89}$ Zhengzhou University, Zhengzhou 450001, People's Republic of China\\

\vspace{0.2cm}
$^{\dagger}$ Deceased\\
% ...
$^{c}$ Also at the Novosibirsk State University, Novosibirsk, 630090, Russia\\
% ...

}

\end{sloppypar}
\end{document}